\documentclass[10pt,journal,comsoc]{IEEEtran}

\usepackage{graphicx}
\graphicspath{{./figures/}}
\DeclareGraphicsExtensions{.eps}

\usepackage{amsmath}
\usepackage{amsfonts}
\usepackage{color}
\usepackage{amssymb}
\usepackage{algorithm}
\usepackage{algorithmic}
\usepackage{array}
\usepackage{multirow}
\usepackage{url}
\usepackage{subfigure}
\usepackage{cite}
\allowdisplaybreaks[1]

\ifCLASSOPTIONcompsoc
  \usepackage[caption=false,font=normalsize,labelfont=sf,textfont=sf]{subfig}
\else
  \usepackage[caption=false,font=footnotesize]{subfig}
\fi

\begin{document}
%
\title{Predictive Switch-Controller Association and Control Devolution for SDN Systems}
%
%
%
%

\author{
Xi~Huang,~\IEEEmembership{Student Member,~IEEE}, 
Simeng~Bian,~\IEEEmembership{Student Member,~IEEE},\\
Ziyu~Shao$^{*}$,~\IEEEmembership{Senior Member,~IEEE},
Hong~Xu,~\IEEEmembership{Senior Member,~IEEE}
\thanks{
{\indent \!\!\!\!
$^*$ The corresponding author of this work is Ziyu Shao.}}
\thanks{
{\indent X. Huang, S. Bian, and Z. Shao are with the School of Information Science and Technology, ShanghaiTech University, China. (E-mail:\{huangxi, biansm, shaozy\}@shanghaitech.edu.cn)
		
	\,\, H. Xu is with Department of Computer Science, City University of Hong Kong, Hong Kong. (E-mail: henry.xu@cityu.edu.hk)}
}%
}

\IEEEtitleabstractindextext{%
\begin{abstract}

For software-defined networking (SDN) systems, to enhance the scalability and reliability of control plane, existing solutions adopt either multi-controller design with static switch-controller association, or static control devolution by delegating certain request processing back to switches. Such solutions can fall short in face of temporal variations of request traffics, incurring considerable local computation costs on switches and their communication costs to controllers. So far, it still remains an open problem to develop a joint online scheme that conducts dynamic switch-controller association and dynamic control devolution. In addition, the fundamental benefits of predictive scheduling to SDN systems still remain unexplored. In this paper, we identify the non-trivial trade-off in such a joint design and formulate a stochastic network optimization problem which aims to minimize time-averaged total system costs and ensure long-term queue stability. By exploiting the unique problem structure, we devise a predictive online switch-controller association and control devolution (POSCAD) scheme, which solves the problem through a series of online distributed decision making. Theoretical analysis shows that without prediction, POSCAD can achieve near-optimal total system costs with tunable trade-off for queue stability. 
With prediction, POSCAD can achieve even better performance with shorter latencies. We conduct extensive simulations to evaluate POSCAD. Notably, with mild-value of future information, POSCAD incurs a significant reduction in request latencies, even when faced with prediction errors. 
\end{abstract}

\begin{IEEEkeywords}
SDN, switch-controller association, control devolution, predictive scheduling.
\end{IEEEkeywords}}

\maketitle

\IEEEdisplaynontitleabstractindextext

%
\IEEEpeerreviewmaketitle

\section{Introduction}\label{sec:introduction}
\IEEEPARstart{D}{uring} the past decade, software-defined networking (SDN) has initiated a profound revolution in networking system design towards more efficient network management and more flexible network programmability. The key idea of SDN is to separate the control plane from the data plane \cite{mckeown2008openflow}. In such a way, the control plane maintains network-wide information to process requests that are constantly generated from SDN-enabled switches in the data plane. Based on the instructions from the control plane, the data plane only needs to carry out basic network functions such as monitoring and packet forwarding.

As the scale of data plane expands, scalability and reliability of control plane often become the main concerns for SDN systems \cite{kreutz2014software}. 
For example, the control plane, if implemented with a singleton controller, can be overloaded by the ever-increasing request traffic, leading to excessively long queueing latencies and belated response to network events. Besides, the singleton controller is also a single point of failure which can result in the breakdown of the whole networking system.

To address such concerns, existing solutions basically fall into two categories. 
One is to implement the control plane as a distributed system with multiple controllers \cite{bannour2018distributed}. Such controllers cooperate to form a logically centralized control plane to manage the network within a single administrative domain \cite{berde2014onos} \cite{koponen2010onix}, or logically distributed control planes to handle networks across different domains \cite{phemius2014disco}\cite{santos2014decentralizing}. 
Under such designs, each switch may have potential connections to multiple controllers for fault tolerance and load balancing. 
Accordingly, the associations between switches and controllers need to be carefully determined so as to reduce the communication costs and balance the workloads among controllers. 
To this end, existing works \cite{dixit2013towards}\cite{filali2018sdn}\cite{krishnamurthy2014pratyaastha}\cite{levin2012logically}\cite{lyu2018multi}\cite{wang2016dynamic}\cite{wang2017efficient} have proposed various solutions to decide \textit{switch-controller associations} in a deterministic fashion. 
The other is to \textit{devolve} the processing of some requests that requires no global information onto switches, to reduce workloads on the control plane \cite{hassas2012kandoo}\cite{zheng2015lazyctrl}. Such techniques have been widely considered and adopted in various large-scale systems such as data center networks \cite{curtis2011devoflow}, WAN \cite{jain2013b4}, and edge computing \cite{xu2016sdn}. For example, Curtis \textit{et al.} \cite{curtis2011devoflow} modified the OpenFlow model to conduct effective flow management by: 1) devolving the control of most flows back onto switches and processing them with an aggressive use of flow-match wildcards or hash-based routing, while 2) controllers maintain global visibility to handle only targeted significant flows; \textit{e.g.}, carrying out load balancing for long-lived flows with high throughputs.

Based on such investigations, we identify several interesting but unresolved questions regarding the control plane design:
\begin{itemize}
  \item[$\diamond$] Instead of conducting deterministic switch-controller associations with infrequent re-association \cite{krishnamurthy2014pratyaastha} \cite{wang2016dynamic}, can we directly perform dynamic association with respect to request traffic variations? What is the benefit of fine-grained control at the request level?
  \item[$\diamond$] How to conduct dynamic devolution?
  \item[$\diamond$] Are there any trade-offs in the joint design of dynamic switch-controller association and dynamic control devolution? If so, how do we manage such trade-offs?   
  \item[$\diamond$] Since the uncertainty in request traffic statistics remains one of the key factors that bring challenges to SDN system design, then if they can be learned, what are the \textit{fundamental} limits of the benefits of attaining such future information for SDN systems?
\end{itemize}

Notably, the last question is motivated by the recent growing interests in leveraging machine-learning-based predictive analytics and scheduling to improve system performance, such as traffic prediction for routing optimization \cite{xie2018survey}, quality-of-experience (QoE) prediction to promote user satisfaction \cite{NetflixPred}, and failure prediction to optimize IT operations \cite{masood2019aiops}. 
Despite the proposal of various prediction-based approaches \cite{chen2016using} \cite{nanda2016predicting} \cite{yu2016power}\cite{zhang2016proactive} in recent years, the \textit{fundamental} benefits of predictive scheduling for SDN systems remain unexplored.

In this paper, we focus on general SDN systems with requests dynamically generated from switches in the data plane for the processing of various network events. We assume that each request can be either processed at a switch (with computation costs) or be uploaded to certain controllers (with communication costs).\footnote{The scenario in which some requests can only be processed by a controller is a special case of our model.} 
We aim to reduce the computational costs by control devolution at data plane, the communication costs by switch-user association between data plane and control plane, and the response times experienced by {switches' requests}, 
which is mainly caused by queueing delays on controllers.
Regarding predictive scheduling, switches are assumed able to predict requests to arrive in a limited number of time slots ahead 
through lightweight prediction modules with recent time-series forecasting techniques\cite{box2015time}.
Further, we assume such future requests can be generated and pre-served before their arrival,
and, if mis-predicted, they will incur extra system costs of communication and computation.\footnote{The scenario in which some future requests may not be pre-served due to their dependency on the processing of previous results is also a special case of our model.} 
Under such settings, we open up a new perspective to answer the above questions. We summarize our contributions as follows.
\begin{itemize}
	\item[$\diamond$] \textbf{Modeling and Formulation:} 
		We formulate the problem stated above as a stochastic optimization problem that aims to minimize time-averaged total system costs with long-term queue stability constraints. Through a careful choice in the granularity of modeling and decision making, \textit{i.e.}, to characterize system dynamics at request level and request scheduling on a time-slot basis, we achieve a decent balance between modeling accuracy and decision making complexity.
	\item[$\diamond$] \textbf{Algorithm Design:} By adopting existing techniques \cite{neely2010stochastic}\cite{huang2016backpressure} and exploiting unique problem structure, we propose \textit{POSCAD}, a Predictive Online Switch-Controller Association and control Devolution scheme which exploits predicted request arrival information to make association and devolution decisions in a distributed fashion.
	\item[$\diamond$] \textbf{Performance Analysis:} We conduct theoretical analysis which shows that without prediction, \textit{POSCAD} yields a tunable trade-off between $O(1/V)$ deviation from minimum long-term average total costs of communication and local computation on switches, and $O(V)$ bound for long-term average queue backlog size. Furthermore, with prediction, \textit{POSCAD} can achieve an even better performance with a notable reduction in request latencies, which is verified by our simulation results. Besides, we also discuss the insights of our design and implementation issues in practice. 
	\item[$\diamond$] \textbf{{Performance Evaluation and Verification:}}\footnote{
			{Note that testbed-based experimental verifications are not considered in this work, since our focus is on the exploration of the \textit{fundamental} limits of the benefits of predictive scheduling in SDN systems. Nonetheless, it can be an interesting direction for future work.}
		} 
	We conduct extensive simulations to evaluate the performance of \textit{POSCAD}. 
	Our results show that under various settings, \textit{POSCAD} achieves near-optimal system costs while maintaining a tunable trade-off with queue stability. Furthermore, given only mild-value of predicted information, \textit{POSCAD} incurs a significant reduction in request response time, even when faced with mis-prediction. 
	\item[$\diamond$] \textbf{New Degree of Freedom in the Design Space of SDN Systems}: 
	To the best of our knowledge, this paper provides the first design to explore and exploit the benefits of predictive scheduling for SDN systems, opening up a new perspective in the design of SDN systems.
\end{itemize}
 
The rest of this paper is organized as follows. 
Section \ref{sec: motivating examples} presents two motivating examples to illustrate the non-trivial trade-off between different performance metrics and the potential benefits of predictive scheduling in SDN systems, respectively.  
Section \ref{sec: problem formulation} demonstrates our system model and problem formulation. 
Then Section \ref{sec: design of poscad} shows the design of \textit{POSCAD} and its performance analysis. 
Section \ref{Section: simulation} discusses our simulation results, Section \ref{sec: related work} reviews related works, while Section \ref{sec: conclusion}  concludes this paper.

\begin{figure}[!t]
\centering
 {
 \includegraphics[width=0.95\columnwidth]{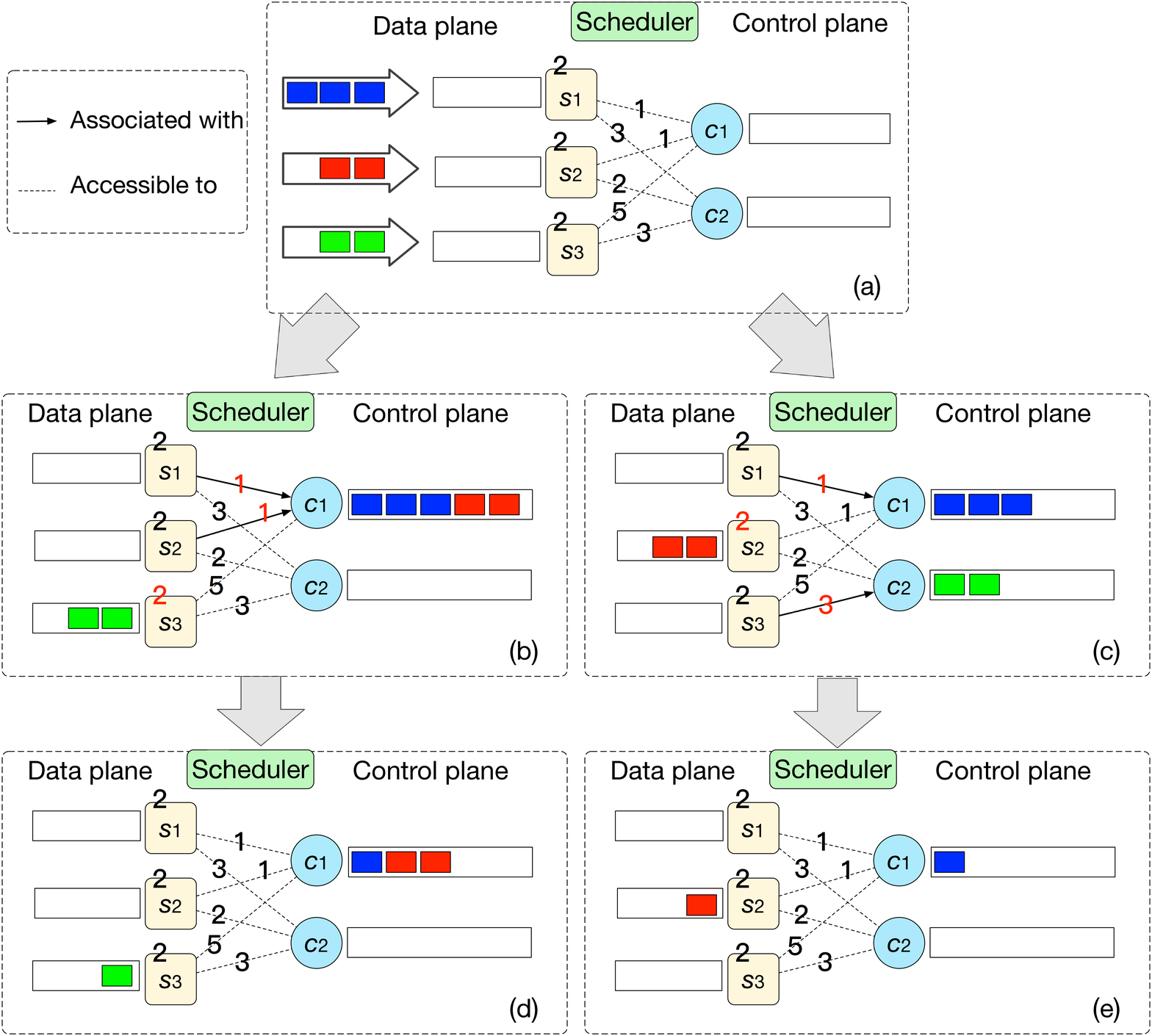}
 }
 \caption{
\textit{An SDN system with request-level switch-controller association and control devolution within one time slot}. 
There are three switches $(s_1,s_2,s_3)$, two controllers $(c_1,c_2)$, and one global scheduler. 
Switches and controllers maintain queue backlogs to buffer requests. Between each pair of controller and switch, each dotted line denotes a potential connection, while each solid line denotes the actual association made in the current time slot. Each connection and each switch are associated with a number denoting the unit communication or local computation costs (on switches) for transferring and processing one request, respectively.
Each controller can serve up to $2$ requests for each time slot, while each switch serves only $1$ request. The scheduler's goal is to minimize the sum of total communication costs and computational costs, as well as the total queue backlog size, which implies timely processing of requests. The system proceeds as follows.
At the beginning of time slot $t$, switches $s_1$, $s_2$, and $s_3$ generate $3$, $2$, and $2$ requests, respectively. 
The scheduler then associates switches to controllers in two ways ((b) and (c)). Switches then process new requests locally or upload them to controllers.}
 \label{fig_sim}
\end{figure}

\section{Motivating Examples}  \label{sec: motivating examples}
In this section, we provide two motivating examples to illustrate the non-trivial trade-off in the joint design of dynamic switch-controller association and control devolution, and the potential benefits of predictive scheduling in SDN systems, respectively.
Such examples motivate our subsequent problem formulation of joint switch-controller association and control devolution with predictive scheduling in SDN systems.

\subsection{Motivating Example of Performance Trade-off}

Figure \ref{fig_sim} shows the evolution of an SDN system within one time slot. 
Particularly, 
Figure \ref{fig_sim} (a) presents the initial system state at the beginning of the time slot.
Figures \ref{fig_sim} (b) -- \ref{fig_sim} (e) show the system evolution given two different association decisions.
In Figure \ref{fig_sim} (b), switches $s_1$ and $s_2$ are associated to controller $c_1$, whereas $s_3$ chooses to process its requests locally. 
In Figure \ref{fig_sim} (d), switches $s_1$ and $s_3$ are associated to controllers $c_1$ and $c_2$, respectively, while $s_2$ chooses to process its requests locally. 

First, we focus on the consequences of different scheduling decisions for switch $s_3$. 
In Figure \ref{fig_sim} (b), 
switch $s_3$ chooses to process its requests locally, which incurs $2$ units of computation costs per request. 
In Figure \ref{fig_sim} (c), $s_3$ decides to upload requests to $c_2$, which incurs $3$ units of communication costs. 
Though the decision in Figure \ref{fig_sim} (b) leads to a lower total cost of $4$ units than Figure \ref{fig_sim} (c) (of $6$ units), it still leaves one request untreated at the end of the time slot.

Next, we switch to the consequences of such scheduling decisions to the whole system. In Figure \ref{fig_sim} (b), the decision (denoted by $X_1$) includes two switch-controller associations: 
$(s_1, c_1)$ and $(s_2, c_1)$ ($s_3$ processes requests locally). From Figure \ref{fig_sim} (d), we see that although decision $X_1$ only incurs total costs of $9$, it still leaves four requests unfinished at the end of the time slot. Meanwhile, in Figure \ref{fig_sim} (c), the decision (denoted by $X_2$) includes two associations: $(s_1, c_1)$ and $(s_3, c_2)$ ($s_2$ processes its requests locally). Figure \ref{fig_sim} (e) show that $X_2$ does better than $X_1$ with two more finished requests at the end of the time slot. However, this is achieved at a higher total cost of $13$ units. 

The above observations show that there is a non-trivial trade-off between system costs and total queue backlog reduction.

 \begin{figure}[!t]
    \begin{center}
        \includegraphics[scale=.225]{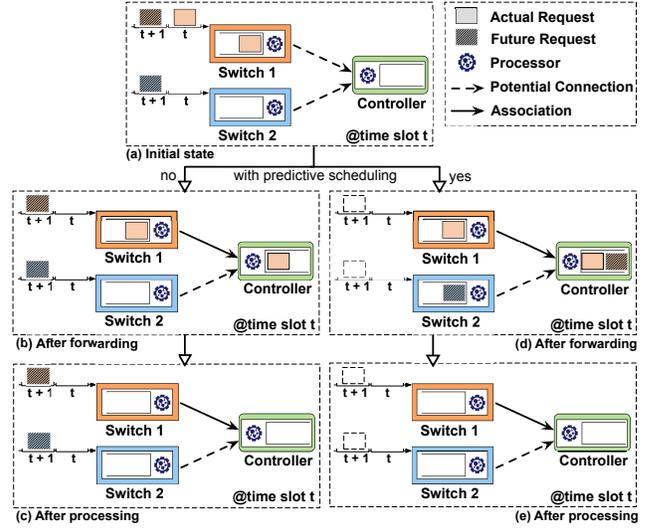}
    \end{center}
    \caption{
    	An example that shows the benefits of predictive scheduling. 
    }
    \label{motiv-example}
\end{figure}

\subsection{Motivating Example of Benefits of Predictive Scheduling} 
Figure \ref{motiv-example} presents an example that compares the cases with and without predictive scheduling, where Figure \ref{motiv-example} (a) shows the system state at the beginning of time slot $t$.
There are two switches potentially connected to one controller (denoted by dashed arrows). 
    	All requests are assumed homogeneous such that they can be handled by both switches and the controller. 
    	Upon new requests' arrival, each switch either associates with the controller (denoted by a solid arrow) and uploads requests, 
    	or stores them in its own local processing queue.
    	In each time slot, a switch processes at most $1$ request and the controller processes at most $2$ requests, both in a first-in-first-out manner. 
    	The objective is to minimize the average request response time. 
    	Figure \ref{motiv-example} (a) shows the system state at the beginning of time slot $t$, where one new request arrives at switch $1$, one is already stored in switch $1$'s local queue, and no requests arrive at switch $2$. 
    	The future request arrivals (marked with stripped colors) in time slot $(t+1)$ are also visible to switches. Figures \ref{motiv-example} (b) and \ref{motiv-example} (c) present the scheduling process that handles current requests only, 
    	whereas Figures \ref{motiv-example} (d) and \ref{motiv-example} (e) exhibit the case with predictive scheduling.

First, we consider the case without predictive scheduling and handle only the arriving requests, as shown in Figures \ref{motiv-example} (b) and \ref{motiv-example} (c).
For switch $1$, it chooses to make an association with the controller and uploads the new request, considering that its local queue has already buffered one request. 
Meanwhile, switch $2$ takes no action since no requests arrive at present. 
After the only new request is forwarded, as shown in Figure \ref{motiv-example} (b), the controller and switches process the requests in their respective queues. 
Figure \ref{motiv-example} (c) shows the system state at the end of time slot $t$, where the two requests in time slot $(t+1)$ are still left unprocessed. 

Next, we focus on the case with predictive scheduling, as shown in Figures \ref{motiv-example} (d) and \ref{motiv-example} (e).
Switch $1$ associates with the controller and uploads the new request that arrives at present. 
Considering that the controller has a service capacity of two requests per time slot, 
it pre-admits the request which will arrive in time slot $(t+1)$, then uploads the request to the controller.  
Similarly, switch $2$ pre-admits the future request in time slot $(t+1)$ and stores the request in its local queue. 
Figure \ref{motiv-example} (d) shows the system state after the (pre-)admission of requests. 
Then the controller and switches consume the requests from their queues. 
Figure \ref{motiv-example} (e) shows that with predictive scheduling, 
all requests in time slots $t$ and $(t+1)$ are completed by the end of time slot $t$. 
Consequently, both future requests will receive instant response upon their arrivals in time slot $(t+1)$. 

The above example shows that predictive scheduling can effectively reduce the request response time by taking advantage of predicted future information.

\vspace{-1.6em}
\begin{center}
\begin{table}[!h]
\caption{Key Notations} \label{notations} 
{
\begin{tabular}{|c|l|}
\hline
\bf{Symbol} & \bf{Description} \\
\hline
\hline
$\mathcal{C}$ & The set of controllers in the control plane \\
\hline
$\mathcal{S}$ & The set of switches in the data plane \\
\hline
$Q^s_i(t)$ & Switch $i$'s local queue backlog size in time slot $t$ \\
\hline
$Q^c_j(t)$ & Controller $j$'s queue backlog in time slot $t$ \\
\hline
\multirow{2}*{$Q^p_i(t)$} & Prediction queue backlog with respect to switch $i$ in \\
& time slot $t$ \\
\hline
\multirow{2}*{$Q^{(d)}_i(t)$} & The number of untreated requests by time slot $t$ \\ 
& that will arrive at switch $i$ in time slot $(t+d)$ \\
\hline
$A_i(t)$ & The number of request arrivals on switch $i$ in time slot $t$ \\
\hline
$B^{s}_i(t)$ & The service capacity of switch $i$ in time slot $t$ \\
\hline
$B^{c}_j(t)$ & The service capacity of controller $j$ in time slot $t$ \\
\hline
\multirow{2}*{$M_{i, j}(t)$} & The per-request communication cost between switch $i$ \\
& and controller $j$ in time slot $t$ \\ 
\hline
\multirow{2}*{$P_{i}(t)$} & The per-request computational cost on switch $i$ in time \\
& slot $t$ \\
\hline
\multirow{2}*{$X_{i,j}(t)$} & The association decision with respect to switch $i$ and \\
& controller $j$ in time slot $t$ \\
\hline
{$Y_{i}(t)$} & The admission decision for switch $i$ in time slot $t$ \\
\hline
{$f_{p}(t)$} & The total communication costs incurred in time slot $t$ \\
\hline
{$g_{p}(t)$} & The total computational costs incurred in time slot $t$ \\ 
\hline
{$h_{p}(t)$} & The weighted total queue backlog size in time slot $t$ \\ 
\hline
\end{tabular}
}
\end{table}
\end{center}
\vspace{-2.9em}

\section{Problem Formulation}\label{sec: problem formulation} 

In this section, we present our system model and problem formulation, with key notations summarized in Table \ref{notations}.
\vspace{-0.3em}

\subsection{System Model} \label{Subsection: formulation without predictive scheduling}

\subsubsection{Basic Model}
We consider an SDN system that evolves over time slots, indexed by $t \in \{ 0, 1, 2, \dots \}$. The system maintains a logically centralized control plane that comprises a set $\mathcal{C}$ of physically distributed controllers. 
Meanwhile, its data plane consists of a group of SDN-enabled switches $\mathcal{S}$. Each switch $i$ keeps a local processing queue\footnote{
		In some other designs \cite{hassas2012kandoo}, the control devolution can also be implemented with a two-layered control plane. Controllers in the bottom layer are deployed in the virtual machines on the same servers as some Open vSwitches. Such controllers have no network-wide information but they can provide faster processing of some basic functions such as local load balancing. Our model can be directly extended to handle such cases by regarding the local processing as the service provided by the controllers in the bottom layer.} of backlog size $Q^{s}_{i}(t)$, while each controller $j$ maintains a queue backlog $Q^c_{j}(t)$ that buffers requests from the data plane. 
	
In each time slot $t$, there are a number of $A_i(t)$ ($\le a_{max}$ for some constant $a_{max}$) new requests arriving at each switch $i$.\footnote{
		{In our work, we define a request as follows. When a new packet arrives at an SDN-enabled switch, the switch will extract the packet’s header fields to match against flow table entries and execute the matched action upon the packet. If the packet matches no entries (or matches some pre-specified entry), the switch will trigger a packet-in event, encapsulate the packet, then forward it to one of the controllers. In this process, we call the processing demand of each packet a \textit{request}.}
	} 
Such arrivals $A_i(t)$ are assumed independently and identically distributed over time slots.
We denote $\mathbf{A}(t) \triangleq \{ A_{i}(t) \}_{i}$. 
{There are two cases of serving a request. One case is local processing, which means that the switch is programmed to provide local control functions to process the packet with sub-optimal but faster decision making\cite{curtis2011devoflow}. The other case is when the packet is added into an SDN event and uploaded to the control plane to decide how to update the flow table.}

In this work, we assume all requests are homogeneous; \textit{i.e.}, they can be handled by both switches and controllers, though our model can be directly extended to scenarios with heterogeneous requests with stateful processing requirements. 
To process such requests, each controller $j$ has a service capacity of $B^{c}_j(t)$ requests, 
while each switch $i \in \mathcal{S}$ has a service capacity of $B^{s}_i(t)$ requests. 
We denote all service rates $\{B^{c}_j(t)\}_{j \in \mathcal{C}}$ and $\{B^{s}_i(t)\}_{i \in \mathcal{S}}$ by $\mathbf{B}(t)$.
Considering the resource limits on switches and controllers, we assume that 
    $B^c_j(t) \le b^c_{max}$ and
    $B^s_i(t) \le b^s_{max}$, for some constants $b^c_{max}$ and $b^{s}_{max}$.
Besides, we also assume the existence of $E\{ \left(A_i(t)\right)^2 \}$, $E\{ ( B^{c}_j(t))^2 \}$, and $E\{ ( B^{s}_i(t) )^2 \}$.

\subsubsection{Pre-service Model} Besides the processing of actually arriving requests, we take a further step by considering the case when the system can predict future request arrivals in a finite number of time slots ahead.\footnote{
		We do not assume any particular prediction techniques in this paper, since our main focus is to explore the fundamental benefits of predictive scheduling. In practice, such prediction can be carried out by applying various machine learning techniques such as time-series prediction methods \cite{box2015time}.} 
Meanwhile, pre-serving future requests is also assumed applicable.\footnote{
	The techniques of request pre-admission are still under active development. Here we take the flow management in SDN as an example. In practice, we can adopt recently developed network traffic prediction techniques \cite{azzouni2018neutm}\cite{liu2016openmeasure}, so that switches and controllers can pre-identify long-lived or bursty flows and pre-install rules on switches to optimize the processing of such flows.}
Particularly, each switch is installed with a learning module that actively predicts request arrivals \cite{xie2018survey}, while maintaining extra buffer for predicted requests. 
Such requests are predicted, pre-generated (with one bit for indication in their headers), and recorded by the switch. Arriving and predicted requests, if scheduled, are appended to corresponding processing queues and later served with some queueing discipline, \textit{e.g.}, first-in-first-out. When predicted requests arrive, if pre-served, they will be considered finished.

Formally, for each time slot $t$, we consider each switch $i$ having access to its future request arrivals in a prediction window of size $D_i$ ($< D$ for some constant $D$), denoted by $\{A_i(t+1), \dots, A_i(t+D_i) \}$. 
In practice, one should leverage machine learning techniques such as time series forecasting methods with low computational complexity \cite{box2015time}, and the size of the prediction window would not be very large, only few time slots. Note that the prediction may be inaccurate often times, leading to extra resource consumptions. We evaluate the impacts of mis-prediction in Section \ref{Section: simulation}.

With predictive scheduling, 
some future requests might have been pre-admitted or pre-served before time slot $t$, thus we use $Q^{(d)}_i(t)$ ($0\leq d\leq D_i$) to denote the number of untreated requests in the $d$-th slot ahead of time $t$, such that 
\begin{equation}
	\begin{array}{c}
		0 \leq Q^{(d)}_i(t) \leq A_i(t+d).
	\end{array}
\end{equation}
Note that $Q^{(0)}_i(t)$ denotes the number of untreated requests that actually arrive at time $t$. 
We denote the total number of untreated requests on each switch $i$ by
$Q^p_i(t) = \sum_{d=0}^{D_i} Q^{(d)}_i(t)$. 
We regard $Q^{p}_i(t)$ as a virtual prediction queue backlog that buffers untreated future requests for switch $i$.

We denote the vector of all queue backlog sizes $\{Q^{p}_i(t)\}_{i \in \mathcal{S}}$, $\{Q^{s}_i(t)\}_{i \in \mathcal{S}}$, and $\{Q^{c}_j(t)\}_{j \in \mathcal{C}}$ by $\mathbf{Q}(t)$.

\begin{figure}[!t]
    \begin{center}
        \includegraphics[scale=1.25]{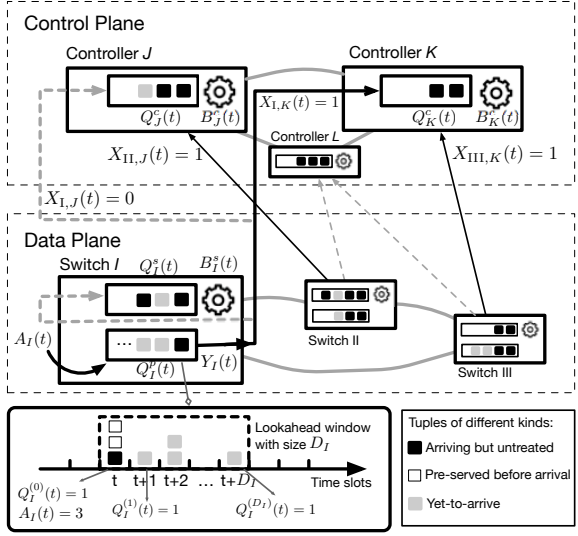}
    \end{center}
    \caption{
        An example of queueing model with predictive scheduling: At the beginning of each time slot, requests arrive at switches over time slots. Along with predicted request arrivals, they are buffered in queue $Q^{p}_{i}(t)$. Meanwhile, each switch $i$ makes association decisions $\{X_{i, j}(t)\}_{j}$ and admission decision $Y_{i}(t)$, then either forwards requests to controllers or processes them locally. Requests are queued up and await to be processed in a first-in-first-out order.}
    \label{queueing-model}
\end{figure}

\subsubsection{Scheduling Decisions}
Upon new requests' arrivals, each switch should make two types of decisions.

Of the first type are \textbf{association decisions}, \textit{i.e.}, deciding to associate with one of its potentially connected controllers or not (\textit{devolution}). 
We adopt $X_{i,j}(t)$ to denote such decisions. Specifically, $X_{i,j}(t) = 1$ if switch $i$ will be associated with controller $j \in \mathcal{C}$ in the current time slot and zero otherwise.\footnote{
		Our model can also be extended to handle the case of logically distributed control planes. Under such designs, each switch belongs to a distinct administrative domain. Accordingly, each switch $i$ only has access to a subset of controllers, which can be denoted by $\mathcal{C}_i \subset \mathcal{C}$.
	} If switch $i$ is assigned with no controllers, \textit{i.e.}, $\sum_{j \in \mathcal{C}} X_{i,j}(t) = 0$, then it appends new requests to its local processing queue.
	An association is feasible if it ensures that each switch is associated with at most one controller during each time slot. The set of feasible associations is defined as follows.
	\begin{equation}
		\mathcal{A} \triangleq \{ 
			\mathbf{X} \in \{0, 1\}^{|\mathcal{S}|\times|\mathcal{C}|}\,\vert\,\sum_{j \in \mathcal{C}} X_{i,j} \leq 1, \forall\,i\in\mathcal{S}
		\}.
	\end{equation}

Of the second type are \textbf{admission decisions}, \textit{i.e.}, deciding the total number of untreated requests to be admitted, including requests that actually arrive in the current time slot and untreated
future requests in its own prediction window.
We use $Y_i(t)$ to denote the number of requests admitted by switch $i$ in time slot $t$.
Such admitted requests must include all untreated requests that have arrived, but not exceed $Q^{p}_i(t)$, \textit{i.e.},
\begin{equation}\label{cst1}
\arraycolsep=1.1pt\def\arraystretch{1.7}
    Q_i^{(0)}(t) \leq Y_i(t) \leq Q_i^{p}(t), ~~\forall i \in \mathcal{S}, t\geq 0.
\end{equation}

In the rest of this paper, we denote the sets $\{X_{i,j}(t)\}_{i\in \mathcal{S}, j\in \mathcal{C}}$ and $\{Y_i(t)\}_{i\in \mathcal{S}}$
by $\mathbf{X}(t)$ and $\mathbf{Y}(t)$, respectively.

\subsubsection{System Workflow and Queueing Model}
Basically, the system proceeds over time slots as follows. At the beginning
of each time slot $t$, new requests arrive at switches. Scheduling decisions $[\mathbf{X}(t),\mathbf{Y}(t)]$ are then made based on instant system dynamics such as $\mathbf{Q}(t)$ and spread to all switches. 
Each switch $i$ then admits $Y_i(t)$ requests, possibly including untreated future requests.
All admitted future requests are marked treated and will not be served again thereafter.
Meanwhile, based upon $\{X_{i,j}(t)\}_{j\in \mathcal{C}}$, each switch $i$ sets up its association and forwards admitted requests accordingly.
Next, all switches and controllers process as many requests as possible from their processing queues. 
At the end of time slot $t$, the learning module on each switch updates its new prediction of request arrivals. 
Accordingly, each switch's prediction window slides ahead by one time slot and new future requests are included to the corresponding prediction queues. 

With the above workflow, we have the following update equations for different queue backlogs. 
\begin{enumerate}
	\item[i)] For the prediction queue with respect to switch $i$, 
	\begin{equation}
	    \label{qupdate_p}
	    \begin{array}{l}
	    \!\!\!\!\!\!\!\!\!\!\!\!\!
	    \displaystyle
    	Q^p_i(t+1) = \left[ Q^p_i(t) - Y_i(t)\right]^{+} + A_i(t+D_i+1).
    	\end{array}
	\end{equation}
	\item[ii)] For the processing queue on switch $i \in \mathcal{S}$, 
	\begin{equation}
	    \label{qupdate_s}
	    \begin{array}{l}
	    	\displaystyle
		    \!\!\!\!\!\!\!\!\!\!\!\!\! Q^s_i(t+1) = 
		    \Big[ Q^s_i(t) + 
		    (1-\sum_{j\in \mathcal{C}} X_{i,j}(t)) Y_i(t) - B_i^s(t)\Big]^{+}.  \\  
	    \end{array}
	\end{equation}
	\item[iii)] For the processing queue on controller $j \in \mathcal{C}$, 
	\begin{equation}\label{qupdate_c}
	\begin{array}{l}
	    \!\!\!\!\!\!\!\!\!\!\!\!\! Q^c_j(t+1) = \left[ Q^c_j(t) + \sum_{i \in \mathcal{S}} X_{i,j}(t) {Y}_{i}(t) - B_j^c(t)\right]^{+}.
	\end{array}
	\end{equation}	
\end{enumerate}
Note that we provide an example of the above queueing model in Figure \ref{queueing-model}.

\subsection{Optimization Objectives}
\textbf{Communication Cost:} 
Request transmissions from data plane to control plane often incur some communication costs.\footnote{In practice, communication costs can be measured by the number of hops, round-trip times (RTT), or transmission powers, \textit{etc.}}
Lower communication costs often imply shorter request response time. For each time slot $t$, we define $M_{i,j}(t)$ as the communication cost of forwarding one request from switch $i$ to controller $j$.
By denoting the set $\{M_{i,j}(t)\}_{i \in \mathcal{S}, j\in\mathcal{C}}$ by $\mathbf{M}(t)$, given decisions $\mathbf{X}(t)$ and $\mathbf{Y}(t)$,
we define the total communication costs in time slot $t$ as
    \begin{equation}\label{def-commcost}
    \begin{split}
		\displaystyle f_{p}(t) &\triangleq \hat{f}_{p} \left( \mathbf{X}(t), \mathbf{Y}(t) \right) = \sum_{i \in \mathcal{S}} \sum_{j \in \mathcal{C}} M_{i,j}(t) X_{i,j}(t) Y_i(t).
	\end{split}
    \end{equation}

\textbf{Computational Cost:} Considering the scarcity of switches' computational resources, the scheduler should spare the use of switches' processing capacities. We use $P_i(t)$ to denote the computational cost for serving a request locally on switch $i$ in time slot $t$. With $\mathbf{P}(t)$ as $\{P_i(t)\}_{i\in\mathcal{S}}$ and given decisions $\mathbf{X}(t)$ and $\mathbf{Y}(t)$, we define the one-time-slot computational costs as
    \begin{equation}\label{def-compcost}
    \begin{split}
		\displaystyle g_{p}(t) &\triangleq \hat{g}_{p}\left( \mathbf{X}(t), \mathbf{Y}(t) \right) \! = \!\!\sum_{i \in \mathcal{S}} P_i(t) \Big[ 1\!-\!\sum_{j\in \mathcal{C}}X_{i,j}(t) \Big] Y_{i}(t).
	\end{split}
    \end{equation}

\textbf{Queue Stability:}
To ensure the timely processing of requests, it is necessary to balance queue backlogs on controllers and switches, so that no queue backlogs would suffer from being overloaded. Hence, we require the stability of all queue backlogs in the system. To this end, we first denote the weighted total queue backlog size in time slot $t$ as
    \begin{equation}
    \arraycolsep=1.1pt\def\arraystretch{1.7}
    \begin{split}\label{def_total_q}
    h_{p}(t) &\triangleq \hat{h}(\mathbf{Q}(t)) = \sum_{j \in \mathcal{C}}\! Q^c_j(t) \!+\! \beta_1 \sum_{i \in \mathcal{S}}\! Q^s_i(t)\! +\! \beta_2 \sum_{i \in \mathcal{S}}\! Q^p_i(t),
    \end{split}
    \end{equation}
    where $\beta_1$ and $\beta_2$ are positive constants that measure the importance of balancing the backlogs of switch queues and prediction queues compared to controller queues, respectively.
	Then we define queue stability \cite{neely2010stochastic} as
    \begin{equation}\label{stability}
    \arraycolsep=1.1pt\def\arraystretch{1.8}
    	\displaystyle 
    		\limsup_{T \to \infty} \frac{1}{T} \sum_{t=0}^{T-1} \mathbb{E}\{ h_p(t) \} < \infty.
    \end{equation}

\subsection{Problem Formulation}
We formulate the following stochastic optimization problem that aims to minimize the long-term time-averaged expectation of the weighted sum of total communication and computation costs, while ensuring long-term queue stability.
\begin{equation}\label{problem1}
\arraycolsep=1.1pt\def\arraystretch{1.7}
	\begin{array}{cl}
	    \underset{\left\{ (\mathbf{X}(t), \mathbf{Y}(t)) \right\}_{t=0}^{T-1}}{\text{Minimize}} &~~ 
	    \displaystyle
	    \limsup_{T \to \infty} \frac{1}{T} \sum_{t=0}^{T-1} \Big[
	    	\mathbb{E}\{ {f}_{p}(t) \} + \gamma \mathbb{E}\{ {g}_{p}(t) \}
	    \Big] \\
        \text{ Subject to } & {X}_{i,j}(t) \in \mathcal{A}, \enskip \forall i \in \mathcal{S}, j \in \mathcal{C}, \\
        & {Y}_i(t) \in \mathbb{Z}_{+}, \enskip \forall i \in \mathcal{S}, 
        \text{ and }
        (\ref{cst1}) - (\ref{qupdate_c}), (\ref{stability}), \\
	\end{array}
\end{equation}
where $\gamma$ is a non-negative constant that measures the scarcity of computation resources on switches.\footnote{
		{
			The overheads of rule installation are not considered in our system model. The reason is that in practice, such overheads are relatively small and can be further mitigated by recently proposed SDN overhead reduction techniques \cite{sanabria2018sdn}. Moreover, our model can also be extended to handle such a case. Specifically, for each switch, we can associate each of its possible choices (including itself and its controllers) with a more general cost that takes such overheads into account, while the rest of our model requires no change.
		}
	}

\section{Algorithm Design and Analysis} \label{sec: design of poscad}

In this section, we present detailed algorithm design to solve problem (\ref{problem1}) with theoretical analysis and further discussion. 

\subsection{Algorithm Design}\label{Subsection: algorithm design for GRAND}

We adopt Lyapunov optimization techniques \cite{neely2010stochastic} to solve problem (\ref{problem1}). First, we define the following Lyapunov function
\begin{equation}\label{lyqueue}
	\begin{array}{cl}
		\displaystyle \!\!\!\! L(\mathbf{Q}(t)) \triangleq \frac{1}{2} \Big[ \sum_{j \in \mathcal{C}} \left(Q^c_j(t)\right)^2 \!\! + \! \beta_1 \sum_{i \in \mathcal{S}} \left(Q^s_i(t)\right)^2 +  
			\beta_2 \sum_{i \in \mathcal{S}} \left(Q^p_i(t)\right)^2
		\Big].
	\end{array}
\end{equation}
Then we define the conditional Lyapunov drift for consecutive time slots as
\begin{equation}\label{cond-drift}
	\begin{array}{c}
		\Delta\left( \mathbf{Q}(t) \right) \triangleq E\left\{ L(\mathbf{Q}(t+1)) - L(\mathbf{Q}(t))\,|\,\mathbf{Q}(t)\right\}.
	\end{array}
\end{equation}
The conditional difference measures the general change in the queueing congestion state of the system. 
Such a difference should be as low as possible to prevent any queue backlog from being overloaded. 
To this end, the decision-making process should aim to minimize (\ref{cond-drift}). However, this may also incur considerable communication and computational costs when faced with various uncertainties in the system, 
such as uneven request arrival traffic on different switches, inhomogeneous service capacities of controllers, and temporal variations of communication costs. 
Hence, besides optimizing (\ref{cond-drift}), such costs should also be considered. Given decisions $\mathbf{X}(t)$ and $\mathbf{Y}(t)$, 
we define the conditional drift-plus-penalty as
\begin{equation}\label{cond-v-drift}
	\begin{array}{c}
		\Delta_V(\mathbf{Q}(t)) \triangleq \Delta(\mathbf{Q}(t)) + V \cdot E\left\{f_{p}(t) + g_{p}(t)|\mathbf{Q}(t)\right\},
	\end{array}
\end{equation}
where parameter $V$ is a positive constant that controls the penalty brought by incurred system costs $f_{p}(t)$ (defined in (\ref{def-commcost})) and $g_{p}(t)$ (defined in (\ref{def-compcost})).
By minimizing the upper bound of the drift-plus-penalty term (\ref{cond-v-drift}), the time-average communication cost can be minimized while stabilizing the network of request queues\cite{neely2010stochastic}. 
We adopt the concept of \emph{opportunistically minimizing an expectation} \cite{neely2010stochastic}, and transform problem (\ref{problem1}) into a series of sub-problems over time slots. Particularly, in each time slot $t$, we have
\begin{equation}\label{problem2}
	\arraycolsep=1.3pt\def\arraystretch{1.7}
	\begin{array}{rl}
		\underset{ \mathbf{X}(t), \mathbf{Y}(t) } {\text{Maximize}} & \displaystyle
		\sum_{i\in \mathcal{S}} l_i(t) Y_i(t) + \sum_{i\in \mathcal{S}} \sum_{j\in \mathcal{C}} u_{i, j}(t) X_{i,j}(t) Y_i(t) \\
    \text{Subject to} & {X}_{i,j}(t) \in \mathcal{A}, \enskip \forall i \in \mathcal{S}, j \in \mathcal{C}, \\
    & {Y}_i(t) \in \mathbb{Z}_{+}, \enskip \forall i \in \mathcal{S}, \text{ and } 
    (\ref{cst1}) -
    (\ref{qupdate_c}),
	\end{array}
\end{equation}
where the terms $l_i(t)$ and $u_{i, j}(t)$ are defined as
\begin{equation}\label{def_l}
    l_i(t) \triangleq \beta_2 Q^p_i(t) - \beta_1 Q_i^s(t) - V\gamma P_i(t),
\end{equation}
\begin{equation}\label{def_w}
    u_{i, j}(t) \triangleq 
    \left[ \beta_1 Q_i^s(t) -  Q_j^c(t) \right] + 
    V \left[ \gamma P_i(t) - M_{i,j}(t) \right],
\end{equation}
respectively.
Note that parameter $V$ is a positive parameter to determine the trade-off between queue stability and the reduction in total system costs. 
The objective function of (\ref{problem2}) can be further decoupled.
For each time slot $t$ and switch $i$:
\begin{equation} \label{problem3}
	\arraycolsep=1.1pt\def\arraystretch{1.6}
	\begin{array}{cl}
		\underset{ \mathbf{X}(t), \mathbf{Y}(t) }
		{\text{Maximize }} & 
		\displaystyle
			l_i(t) \cdot Y_i(t) + 
			\Big[ \sum_{j\in \mathcal{C}} u_{i, j}(t) X_{i,j}(t) \Big] \cdot
			Y_i(t)
			\\
    \text{Subject to } & \displaystyle 
    X_{i, j}(t) \in \mathcal{A}, \enskip 
	Y_{i}(t) \in \mathbb{Z}_{+}, \enskip
    \forall\ j \in \mathcal{C}, \\
    & \displaystyle
    Q^{(0)}_i(t) \le Y_i(t) \le Q^{p}_i(t).
    \end{array}
\end{equation}
Note that variables $\{X_{i, j}(t)\}_{j \in \mathcal{C}}$ and $Y_{i}(t)$ are coupled in the objective function, making problem (\ref{problem3}) complicated to solve.
However, by exploiting the problem structure, we show that it can be solved optimally, as follows. 

Given $[\mathbf{Q}(t), \mathbf{P}(t), \mathbf{M}(t)]$, $l_i(t)$ and $u_{i, j}(t)$ are constants in time slot $t$. 
For each switch $i$, 
solving problem (\ref{problem3}) is equivalent to finding the maximum product of two terms: 1) $Y_{i}(t)$ (positive and bounded), 2) $l_{i}(t) + \sum_{j \in \mathcal{C}} X_{i, j}(t) u_{i, j}(t)$, which depends on $\{X_{i, j}(t)\}_{j \in \mathcal{C}}$ that at most one of them equals one. 

To achieve the maximum value of (\ref{problem3}), we consider the following three cases. 

	\textit{Case 1}: If $l_{i}(t)\!<\!0$ and $l_{i}(t)\! +\! u_{i, j}(t)\!<\!0$ for all $j$, 
	then one should consider two sub-cases. In either case, $Y_{i}(t)$ should be set as its minimum $Q^{(0)}_{i}$, and we aim to decide $\{X_{i, j}(t)\}_{j \in \mathcal{C}}$ that maximizes the value of $l_{i}(t) + \sum_{j \in \mathcal{C}} X_{i, j}(t) u_{i, j}(t)$. 
	\begin{enumerate}
		\item[i)] If $u_{i, j}(t) < 0$ for all $j$, then choose no controllers, \textit{i.e.}, $X_{i, j}(t) = 0$ for all $j$.
		\item[ii)] Otherwise, if $\hat{\mathcal{C}} \triangleq \{ j'\!\in\!\mathcal{C} 
			\ \vert\ 
			u_{i, j'}(t) \geq 0\} \neq \emptyset$, then 
		one should choose the controller $j^{*} \in \arg\max_{j' \in \hat{\mathcal{C}}} u_{i, j'}(t)$ by setting $X_{i, j^*}(t) = 1$ and others to be zero.\footnote{If more than one controller achieves the maximum, then the switch should choose one of them uniformly at random.}
	\end{enumerate}
	
	\textit{Case 2}: If $l_{i}(t) \geq 0$ and $l_{i}(t) + u_{i, j}(t) < 0$ for all $j$, then the best choice is to choose no controllers (so that the second term equals $l_{i}(t)$) and set $Y_{i}(t)$ to its maximum $Q^{p}_{i}(t)$.
	
	\textit{Case 3}: If $\tilde{\mathcal{C}} \triangleq \{ j'\!\in\!\mathcal{C} 
			\ \vert\ l_{i}(t) + u_{i, j}(t) \geq 0 \} \neq \emptyset$,
			then the optimal solution is attained by setting $X_{i, j^{*}}(t) = 1$ for $j^{*} \in \arg\max_{j' \in \tilde{\mathcal{C}}} u_{i, j'}(t)$ and others to be zero, while $Y_{i}(t)=Q^{p}_{i}(t)$.

In this way, problem (\ref{problem3}) can be solved optimally. 
Based the above design, we propose \textit{POSCAD}, a predictive and distributed scheme to solve problem (\ref{problem3}). Its pseudocode is given in Algorithm \ref{algo}. 

\begin{algorithm}[!t]
 \caption{POSCAD (Predictive Online Switch-Controller Association and control Devolution) in one time slot}
 \begin{algorithmic}[1] \label{algo}
 \renewcommand{\algorithmicrequire}{\textbf{Input:}}
 \renewcommand{\algorithmicensure}{\textbf{Output:}}
 \REQUIRE Queue backlog sizes $\mathbf{Q}(t)$, computation costs $\mathbf{P}(t)$ on switches, and communication costs $\mathbf{M}(t)$ in time slot $t$.
 \ENSURE Admission and association decisions $\mathbf{Y}$ and $\mathbf{X}$.
 \\ 
    \STATE \textbf{for} each switch $i \in \mathcal{S}$
  	\STATE $~~$ Calculate $l_i(t)$ and $u_{i, j}(t)$ for each controller $j \in \mathcal{C}$\\
  		  $~~$ according to 
 		  (\ref{def_l}) and (\ref{def_w}), respectively.
      \STATE $~~$ \textbf{if} $u_{i, j}(t) < 0$ for each controller $j \in \mathcal{C}$
      	\textbf{then}
  	  \STATE $~~$ $~~$ \textbf{if} $l_i(t) > 0$ \textbf{then}
  	  \STATE 
  	  $~~$ $~~$ $~~~$
  	  set $Y_i \leftarrow Q^{p}_{i}(t)$.
  	  \STATE $~~$ $~~$ \textbf{else}
  	  set $Y_i \leftarrow Q^{(0)}_{i}(t)$. \\
  	  \STATE $~$ $~$ $~$ \textbf{endif}
  	  \STATE $~$ $~$ $~$
	  Set $X_{i,j} \leftarrow 0$, $\forall j$.
  	  \STATE $~$ $~$ $~$
  	  Admit $Y_i$ requests from its prediction queue and 
  	  \STATE $~$ $~$ $~$
  	  append them to switch $i$'s local queue $Q^{s}_{i}(t)$. 
  	  \STATE $~~$ 
  	  \textbf{else} 
  	  associate switch $i$ with the controller $j^{*}$ such that
  	  \begin{equation}
  	  	\begin{array}{c}
  	  	j^{*} \in \underset{j \in \mathcal{C}}{\arg\max}~u_{i, j}(t). \nonumber	
  	  	\end{array}
  	  \end{equation}
  	  \STATE $~~$ $~~$ \textbf{if} $l_i(t) + u_{i,j^{*}}(t) > 0$ \textbf{then}
  	  \STATE 
  	  $~~$ $~~$ $~~~$
  	  set $Y_i \leftarrow Q^p_i(t)$.
  	  \STATE $~~$ $~~$ \textbf{else}
  	  set $Y_i \leftarrow Q^{(0)}_{i}(t)$. \\
  	  \STATE $~$ $~$ $~$ \textbf{endif}
  	  \STATE $~$ $~$ $~$
  	  Admit $Y_i$ requests from its prediction queue. 
  	  \STATE $~$ $~$ $~$
  	  Forward admitted requests to controller $j^*$ and add \\
  	  $~$ $~$ $~$
  	  them to $Q^{c}_{j^*}(t)$; \textit{i.e.}, set $X_{i,j^*} \leftarrow 1$ and $0$ for $j \neq j^*$.
  	  \STATE $~~$ \textbf{endif}
  	  \STATE \textbf{endfor}
  	  \STATE All switches and controllers then consume requests from their respective processing queues. 
\end{algorithmic}
\end{algorithm}
\setlength{\floatsep}{-3pt}

\subsection{Discussion:}
First, we discuss the roles of $l_i(t)$ and $u_{i, j}(t)$ in \textit{POSCAD}, as defined in (\ref{def_l}) and (\ref{def_w}), respectively.
	On the one hand, the value of $u_{i, j}(t)$ reflects switch $i$'s willingness of associating with controller $j$ in time slot $t$. 
	On the other hand, the value of $l_i(t)$ quantifies the weighted balance between the number of untreated requests in the prediction queue and switch $i$'s local queue backlog size and computation cost. 
	The larger the value of $l_i(t)$, the more untreated future requests in switch $i$'s prediction queue, and the more requests switch $i$ tends to admit.
	Note that both $l_i(t)$ and $u_{i, j}(t)$ utilize queue backlog sizes and estimate communication cost as the indicator of congestion.
	Note that such information is readily available on switches and controllers at the beginning of each time slot.
	
	Based on the above discussion, we illustrate how \textit{POSCAD} works.
	If switch $i$ decides to process new requests locally, \textit{i.e.}, 
	$u_{i, j}(t)\! <\! 0$ for all $j \in \mathcal{C}$, 
	the number of admitted requests depends on the value of $l_{i}(t)$.
	If $l_i(t) > 0$, switch $i$ would admit future requests to its local queue; otherwise, only untreated requests in current time slot will be admitted.
	If $u_{i, j}(t) \! > \! 0$ for some controllers, then switch $i$ will associate with the controller $j^{*}$ with the maximum $u_{i, j^*}(t)$. 
	In this case,  
	\textit{POSCAD} decides request admission based on the value of $l_i(t) + u_{i, j^{*}}(t)$, which by (\ref{def_l}) and (\ref{def_w}) turns out to be
	\begin{equation}\label{def_l_add_w}
		l_i(t) + u_{i, j^{*}}(t) = 
		\beta_2 \cdot Q^p_i(t) - Q_{j^{*}}^c(t) - V \cdot M_{i, j^{*}}(t).
	\end{equation}
	Intuitively, $l_i(t) + u_{i, j^{*}}(t)$ reflects 
	the weighted balance among the number of all untreated requests in the prediction queue, controller $j^{*}$'s queue backlog size, and their communication cost in between.
	Switch $i$ tends to admit more future requests when $l_i(t) + u_{i, j^{*}}(t) > 0$, \textit{i.e.},
	controller $j^{*}$ possesses a queue backlog with a small size and low communication costs. 

Next, we discuss the role of parameter $V$ in \textit{POSCAD}. Specifically, with a sufficiently large value of $V$, the terms related to queue backlogs in (\ref{def_l}) and (\ref{def_w}) become negligible. In this case, \textit{POSCAD} inclines to neglect the workload balance among queue backlogs, and admits only arriving requests by sending them to those queues with the least costs. In contrast, if the value of $V$ approaches zero, \textit{POSCAD} would ignore the impact of such costs and greedily forward requests to the queue with the smallest backlog size.

\subsection{Performance Analysis}
First, we assume that the expectation of total system processing capacity is greater than that of the total request arrival rate. 
Then without predicted information (\textit{i.e.}, $D_i = 0$ for each switch $i \in \mathcal{S}$), we can show that \textit{POSCAD} achieves the classic $[O(V),O(1/V)]$ trade-off between the time-averaged expectation of total costs and total queue backlog size in the system. The proof is quite standard and thus omitted here \cite{neely2010stochastic}. 

Moreover, with predictive scheduling, \textit{POSCAD} can achieve an even better performance by breaking the $[O(V),O(1/V)]$ backlog-cost barrier with even shorter request latencies.
Intuitively, this is achieved by making each switch proactively leverage available predicted information and opportunistically exploit surplus system processing capacities to pre-serve predicted requests during each time slot. Regarding the theoretical analysis for the impacts of prediction errors, it is still an open problem. In this paper, we resort to simulation results to evaluate such impacts. More details are given in Section \ref{Section: simulation}.

Regarding the time complexity of \textit{POSCAD}, from Algorithm \ref{algo}, 
we see that \textit{POSCAD} can be run in a distributed manner. 
In particular, after acquiring instant information system dynamics at the beginning of each time slot, each switch conducts decision making that is independent of each other. 
For each switch, the searching for the best candidate to forward requests requires only a time complexity of $O(|\mathcal{C}|)$. 
Such a low time complexity allows system designers to trade off only little overheads for notable improvements in system performance. More detailed discussion about the implementation of \textit{POSCAD} is given in Section \ref{Section: simulation}.

\begin{figure}[!t]
\centering
  	\vspace{-3.8em}	
 \includegraphics[width=0.45\textwidth]{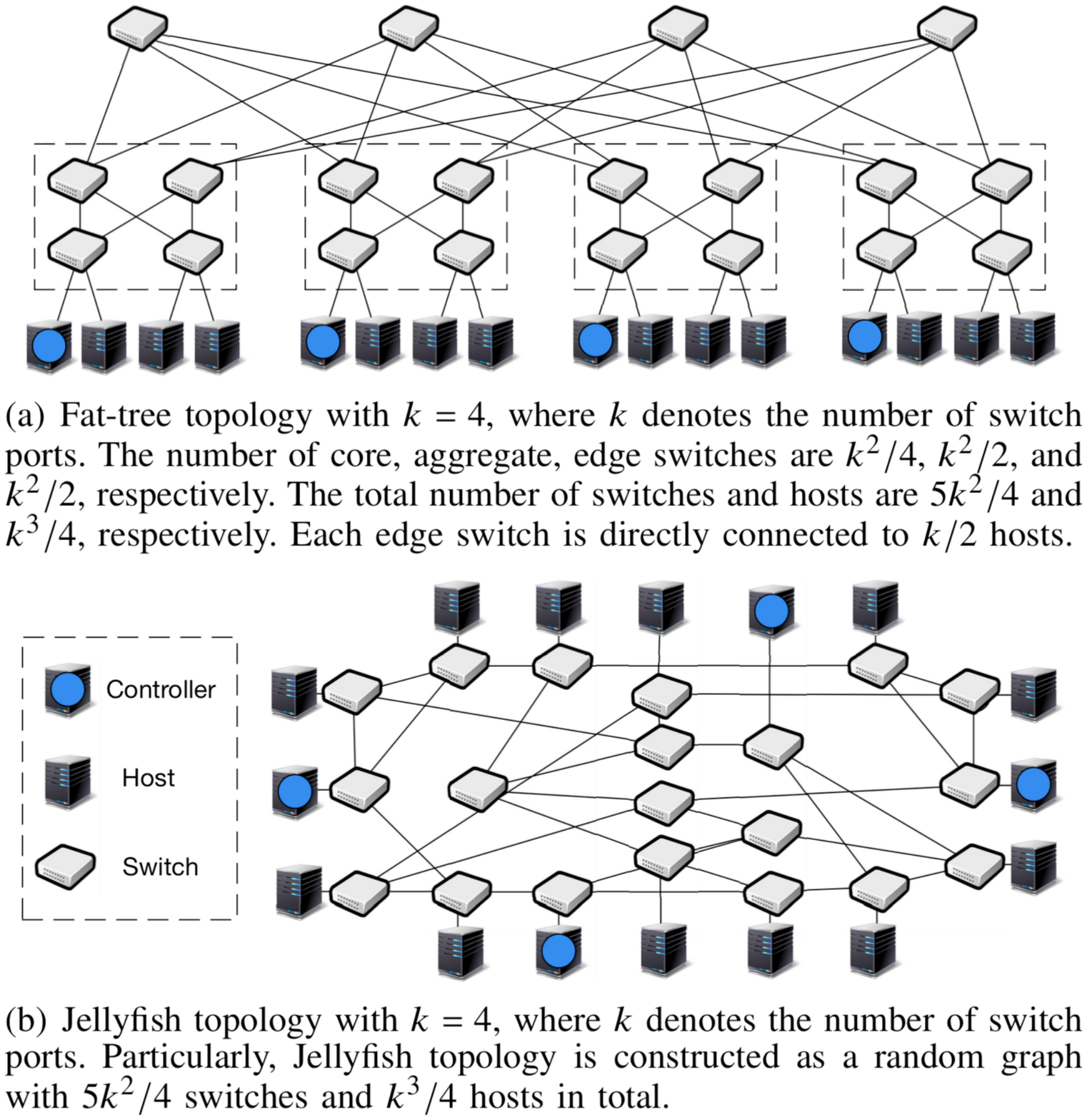}
  	\vspace{-4.3em}	
 \caption{Instances of Fat-tree and Jellyfish topologies for SDN systems. All controllers are deployed on particular hosts.} 
 \vspace{-1em}
 \label{topos}
\end{figure}

\section{Simulation Results}\label{Section: simulation}

We conduct trace-driven simulations to evaluate the performance of \textit{POSCAD} under different settings. 
{Particularly, in the following subsections, we first evaluate the performance of \textit{POSCAD} without prediction (\textit{i.e.}, with prediction window size $D_i = 0$ for each switch $i \in \mathcal{S}$).
For ease of reference, we name such a special case of \textit{POSCAD} as \textit{GOSCAD} (Greedy Online Switch-Controller Association and control Devolution scheme). The reason is that by Algorithm \ref{algo}, when $D_i = 0$ for all $i \in \mathcal{S}$, each switch $i$ always greedily forwards requests to the controller $j$ with the maximum positive value of $u_{i,j}(t)$ (if $u_{i, j}(t) < 0$ for all $j$, then process requests locally).} 
Then we evaluate \textit{POSCAD} under more general scenarios with perfect and imperfect predicted information, respectively.

\subsection{Basic Settings}

\textbf{Topology:} We evaluate the performance of {\textit{POSCAD}} under four well-known data center topologies: Canonical 3-Tiered topology\cite{benson2010network}, Fat-tree\cite{al2008scalable}, Jellyfish\cite{singla2012jellyfish}, and F10\cite{liu2013f10}. We show the instances for Fat-tree and Jellyfish, respectively, in Figure \ref{topos}.
To make our performance analysis fair among the four topologies, we construct them at almost the same scale, comparable to commercial data centers \cite{benson2010network}.

We deploy controllers (the control plane of the SDN system) on particular hosts, which are denoted by hosts with blue circles. Each controller has a service capacity of $600$ requests per time slot (the typical setting of NOX controller \cite{tootoonchian2012controller}).
Specifically, for deterministic topologies (Fat-tree, Canonical 3-Tiered, and F10), we deploy one controller in every pod;\footnote{In Canonical 3-Tiered topology, we regard the group of switches that belong to the same aggregation switch as one pod (including the aggregation switch itself).} 
for random topology (Jellyfish), we deploy the same number of controllers on hosts with non-adjacent ToRs.

\textbf{Request Arrival Settings:} Note that the design of \textit{POSCAD} does not depend on particular traffic statistics. In our simulations, the request inter-arrival times follow the distribution that is drawn from measurements of real-world data centers \cite{benson2010network}. We also consider traffic statistics that follow Poisson and Pareto distributions, since they are widely adopted in queueing network analysis. 
Under such scenarios, the length of each time slot is set as $10$ms and the average request arrival rate on each switch is about $5.88$ requests per time slot. Besides, considering the existence of hot spots ($i.e.$, some switches have intensive request arrivals) in real-world systems, we pick the first pod as the hot spot with each switch having a request arrival rate of $200$ requests per time slot.

\textbf{System Cost Settings:} 
By setting parameter $\gamma = 1$, we assume the equal importance of reducing communication costs and local computational costs on switches. Given any network topology, we define the communication cost $M_{i,j}(t)$ between switch $i$ and controller $j$ as the length (number of hops) of shortest path from $i$ to $j$. 
For each topology, to make switches' computational costs comparable to communication costs, we set the average unit computation cost $P$ of switches as the average hop number between switches and controllers in the topology. In Fat-tree and F10 topologies, $P = 4.13$; in 3-Tiered and Jellyfish topologies, $P$ equals $4.81$ and $3.56$, respectively.

\begin{figure}[!t]
\centering
\vspace{-3.5em}
\includegraphics[width=0.93\columnwidth]{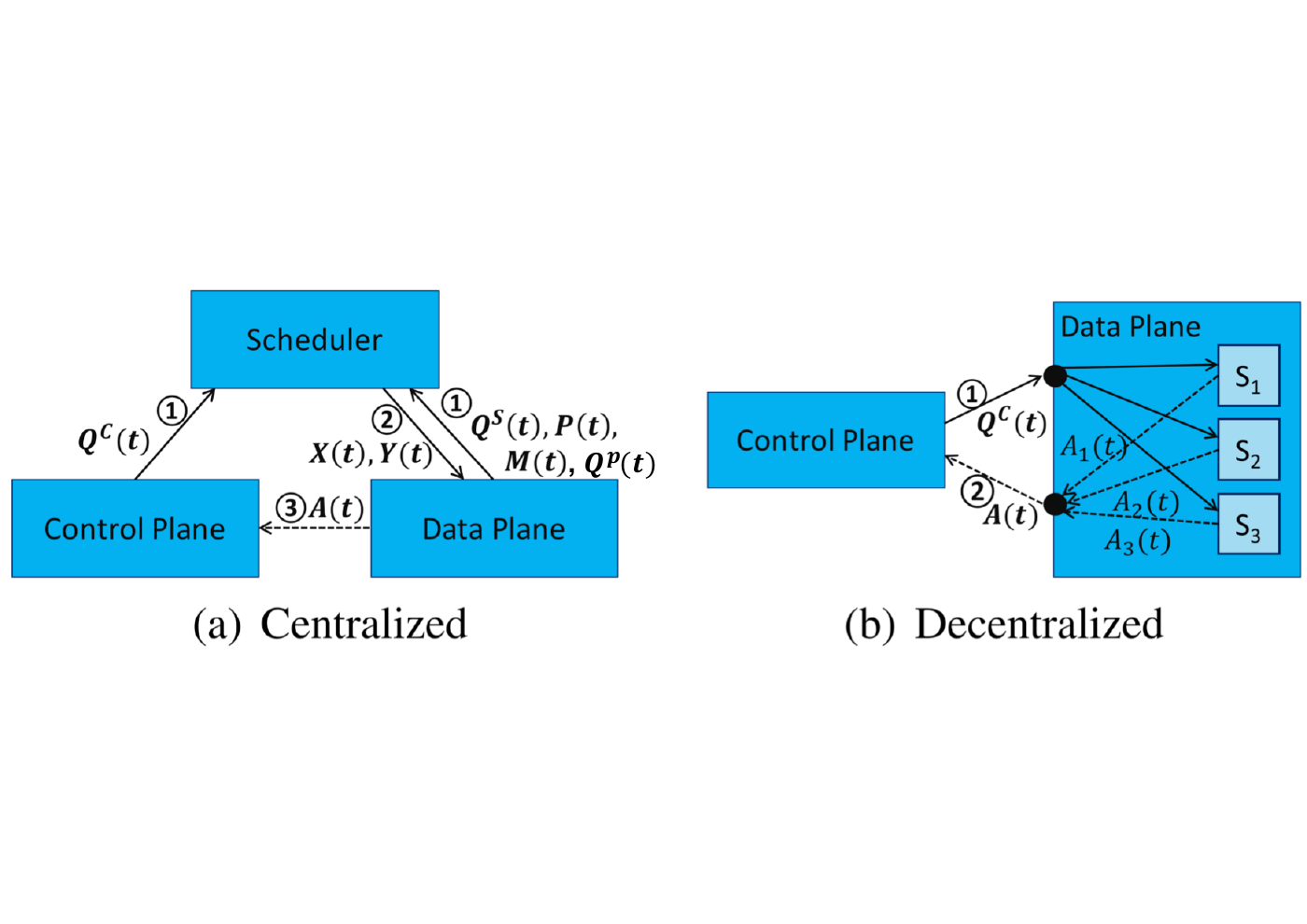}
\vspace{-5em}
 \caption{Two scheduler implementations of \textit{POSCAD}.}
 \vspace{-0.5em}
 \label{architecture}
\end{figure}
	
\textbf{Queueing:}
We treat all queue backlogs equally by setting $\beta_1$ and $\beta_2$ as $1$ in (\ref{def_total_q}).
In practice, the values of parameter $\beta_1$ and $\beta_2$ can be tuned proportional to the ratio between the capacity of prediction queues, switches' queues, and controllers' queues. 
Recall that the value of parameter $V$ determines the importance of communication cost reduction to queue stability. 
When applying \textit{POSCAD}, one can tune the value of parameter $V$ proportional to the ratio between the magnitude of queue backlog size (in number of requests) and communication cost (in milliseconds). 
In our simulation, the ratio is about $100$ and thereby we adjust the value of parameter $V$ from $1$ to $10^4$. 
The queueing policy is {first-in-first-out} (FIFO) for all queues.

\textbf{Scheduler Implementations:} Figures \ref{architecture} (a) and \ref{architecture} (b) illustrate how {\textit{POSCAD}} can be implemented in a centralized and decentralized manner, respectively. Both implementations can be applied to various systems, such as data center networks, SDN-based WANs, and SDN-based edge computing systems.

In the centralized way, the scheduler is independent of both control plane and data plane. The scheduler collects system dynamics including queue backlogs on both switches and controllers to make centralized scheduling decisions. Next, it spreads the scheduling decisions onto switches; then switches upload or locally process their requests according to such decisions. The abstract process is presented in Figure \ref{architecture} (a). The advantage of centralized architecture is that it requires no modifications on data plane, \textit{i.e.}, all the system dynamics such as communication costs and queue backlogs can be obtained via standard OpenFlow APIs. This is well-suited for the situation where the data plane is at a large scale and switch computing resource is scarce. In fact, the scheduler could also be deployed on the control plane. There are disadvantages, too. Centralized scheduler is a potential single point of failure, or even a bottleneck with considerable computation overheads. Besides, it requires back-and-forth message exchange between the SDN system and the scheduler, thus leading to longer latencies.  
{Specifically, under such an implementation, it takes extra bandwidth consumptions and latencies to obtain instant system dynamics (system costs and backlog sizes) from the control plane and the data plane. Nonetheless, the mechanism for probing instant system dynamics can be implemented in a scalable fashion. 
In particular, by deploying \textit{POSCAD} over a group of dedicated servers, each server only needs to probe instant dynamics from part of switches and controllers. Based on probed information, each server can conduct independent decision making for switches.}

In the decentralized way, as Figure \ref{architecture} (b) shows, switches periodically update their information about queue backlogs from control plane; meanwhile, each of them can make scheduling decisions independently. Though this way requires modification on switches, the decentralized way still has the following advantages. It requires less amounts of message exchange than that in the centralized way, thus switches would response even faster to handling network events. On the other hand, the computation of decision-making in both schemes is distributed across switches, leading to better scalability and fault tolerance. In our simulations, we choose the decentralized implementation for \textit{POSCAD}.

\subsection{Evaluation of {\textit{GOSCAD}} (\textit{POSCAD} without Prediction)}

\begin{figure}[!t]
\centering
 \subfigure[Total cost vs. V] {
 \includegraphics[width=0.465\columnwidth]{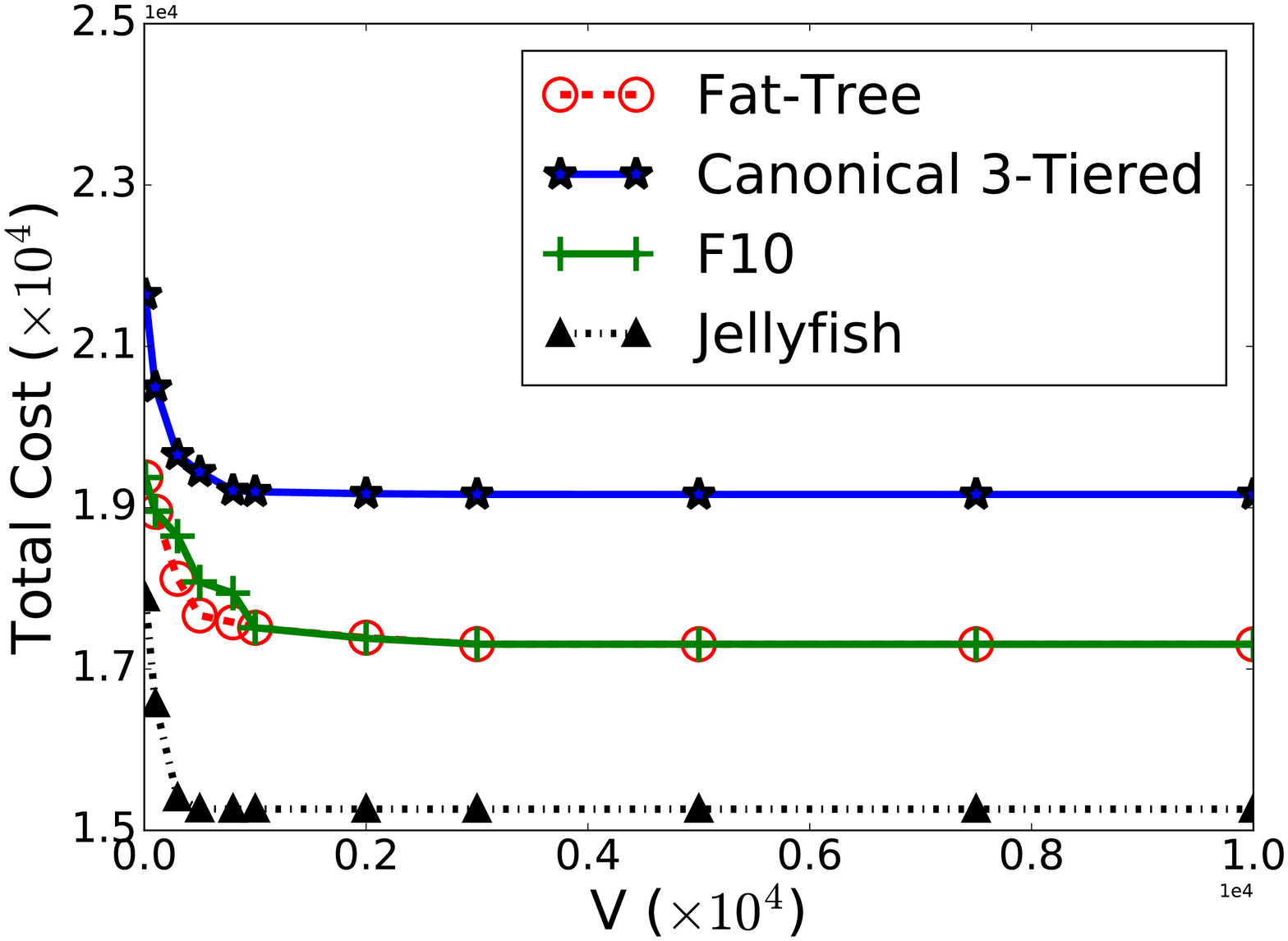}
 }
 \subfigure[Total queue backlog vs. V] {
 \includegraphics[width=0.465\columnwidth]{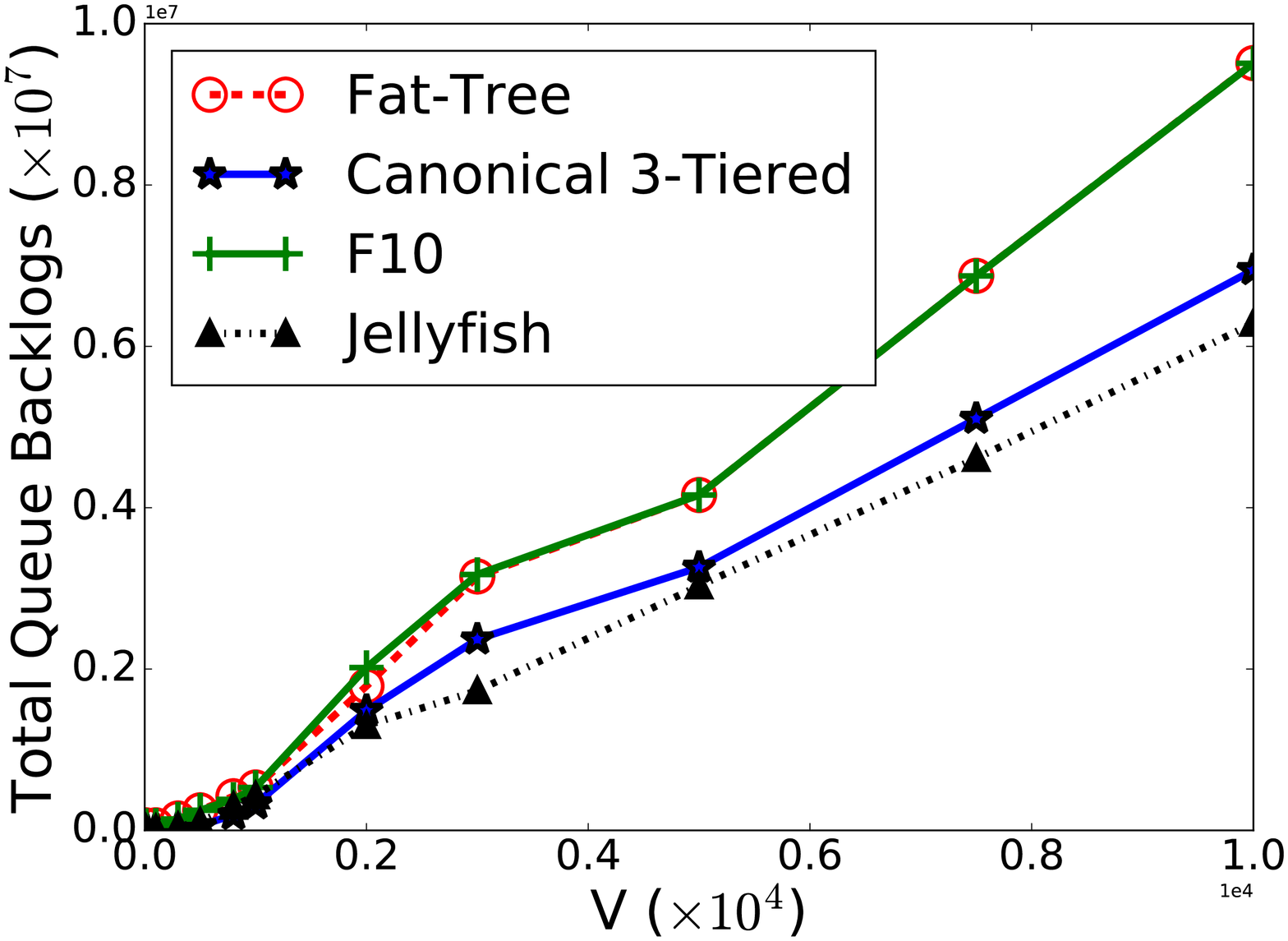}
 }
 \caption{Performance of \textit{GOSCAD} (\textit{POSCAD} without prediction) under different topologies.}
 \label{trace}
\end{figure}

Figures \ref{trace} (a) and \ref{trace} (b) show how different values of parameter $V$ affect the long-term time-averaged total costs of communication and computation, and the total queue backlog size in the system. In general, as the value of parameter $V$ grows from $1$ to $10^4$, we see a rapid reduction in total system costs and a linear increase in the total queue backlog size. However, as the value of $V$ continues to increase, the reduction in total system costs diminishes, while the queue backlog size keeps going up. Note that this verifies our previous theoretic results on the $[O(V),O(1/V)]$ trade-off between the time-averaged expectation of total costs and total queue backlog size in the system. We discuss them in detail as follows.

First, recall that the value of parameter $V$ actually controls the switches' willingness of uploading requests. 
For switches that are close to controllers (their communication costs are less than the average), large values of $V$ make them incline to uploading requests, unless all controllers become heavily loaded. 
For switches that are distant from controllers, local processing is a better choice. Consequently, large values of $V$ generally lead to lower communication costs. However, as the value of $V$ becomes sufficiently large, switches either greedily process requests locally or upload them to the nearest controllers, leading to hot spots on particular controllers with ever-increasing backlog sizes. In this case, according to Little's theorem \cite{little1961proof}, request response time increases as well.

\begin{figure}[!t]
	\centering
	\subfigure[Total cost vs. V] {
		\includegraphics[width=0.45\columnwidth]{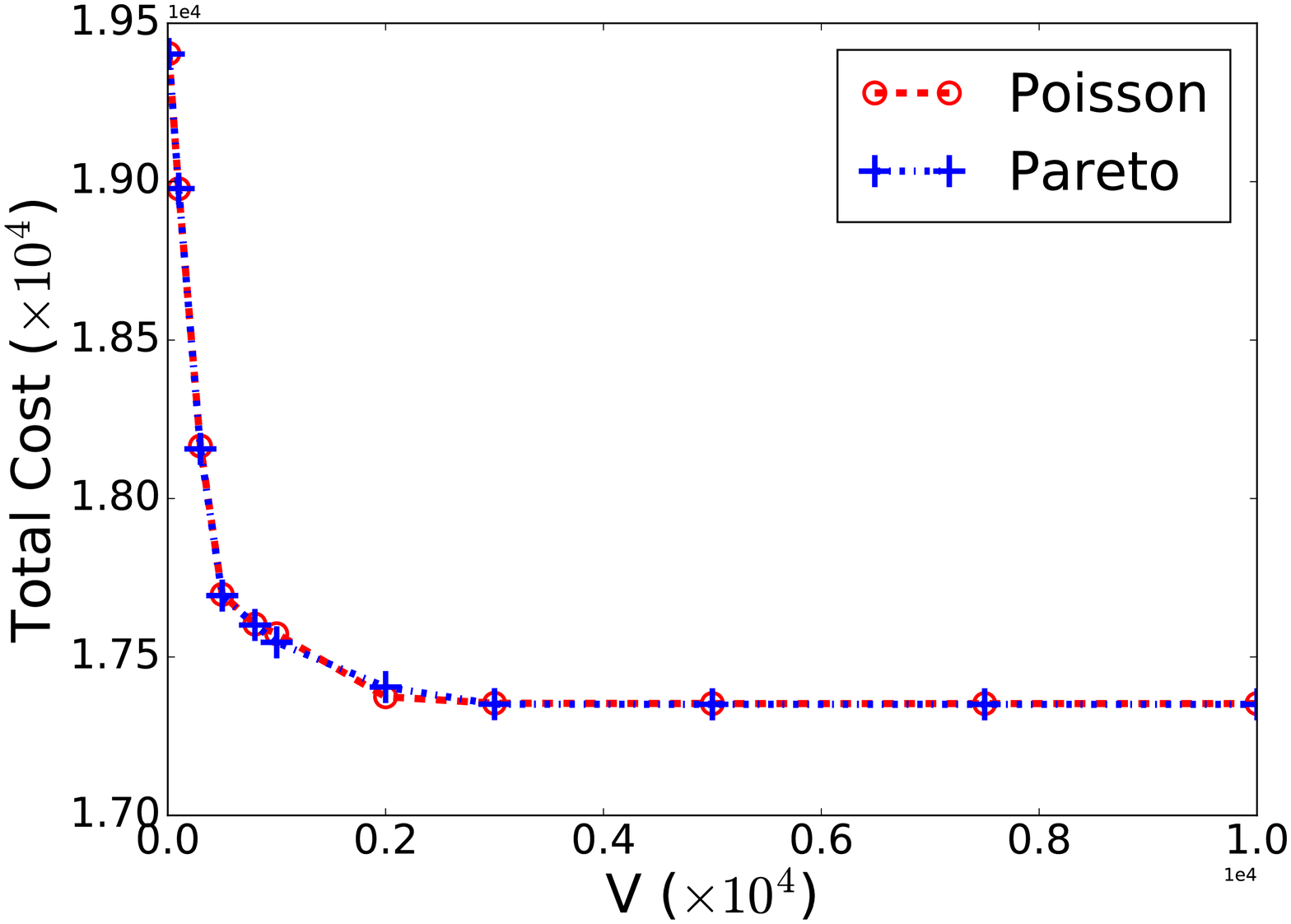}
	}	
	\subfigure[Total queue backlog vs. V] {
		\includegraphics[width=0.45\columnwidth]{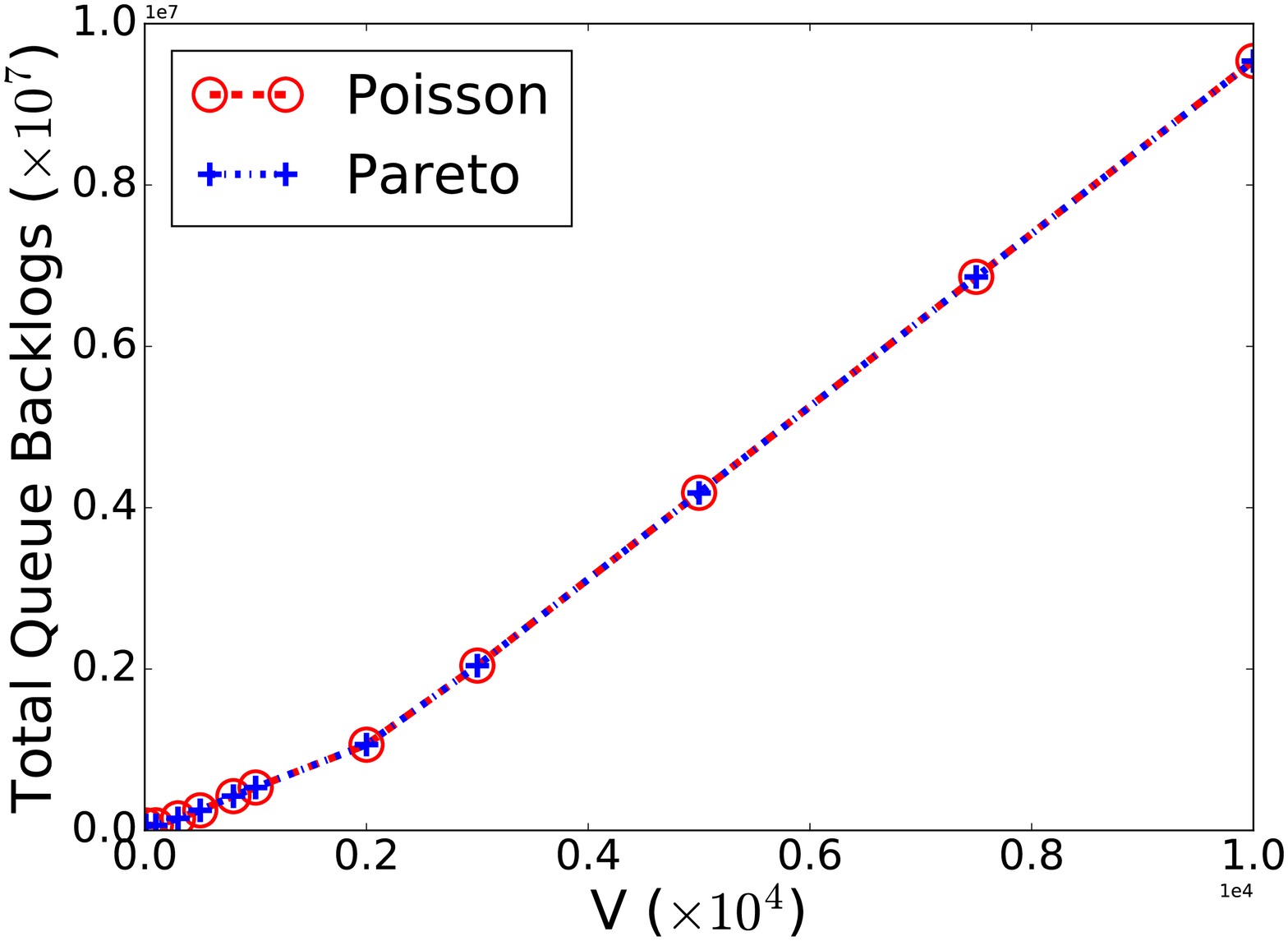}
	}
	\caption{Performance of \textit{GOSCAD} (\textit{POSCAD} without prediction) under Fat-tree topology with other two arrival processes.}
	\label{pp}
\end{figure}
\setlength{\floatsep}{0pt}

Figure \ref{trace} (a) presents how the long-term time-averaged total costs of communication and local computation (on switches) with different values of $V$ in the four topologies. We make the following observations. 

First, as the value of $V$ varies from $1$ to $10^4$, it shows that the total cost goes down gradually. This is consistent with our previous theoretic analysis. The intuition behind such decline is as follows. Remind that $V$ controls the switches' willingness of uploading requests. For switches that are close to controllers (their communication cost is less than the average), large $V$ makes them unwilling to process requests locally unless the controllers are heavily loaded. As the value of $V$ increases, those switches will choose to upload requests to further reduce the costs since for those switches, communication costs are less than the computation costs.

Second, the total cost in 3-Tiered topology is greater than the other schemes', because it has a higher unit computational cost ($P = 4.81$ compared to $4.13$ and $3.56$ for Fat-tree/F10 and Jellyfish). 
Meanwhile, switches in 3-Tiered topology usually take longer paths to controllers and incur more communication costs compared to other topologies.

Third, the total costs under Jellyfish topology are significantly lower (up to $21\%$) than the others, as Jellyfish generally has a shorter average path between switches and controllers than the other deterministic topologies of the same scale \cite{singla2012jellyfish}.

Figures \ref{pp} (a) and \ref{pp} (b) show the performance of \textit{GOSCAD} under Fat-tree topology with other two request arrival processes. The results show that \textit{GOSCAD} (\textit{i.e.}, \textit{POSCAD} without prediction) requires no prior knowledge about the statistics of request arrival dynamics and still achieves the $[O(V),O(1/V)]$ backlog-cost trade-off.

\begin{figure}[!t]
\centering
 \subfigure[Canonical 3-Tiered topology] {
 \includegraphics[width=0.43\columnwidth]{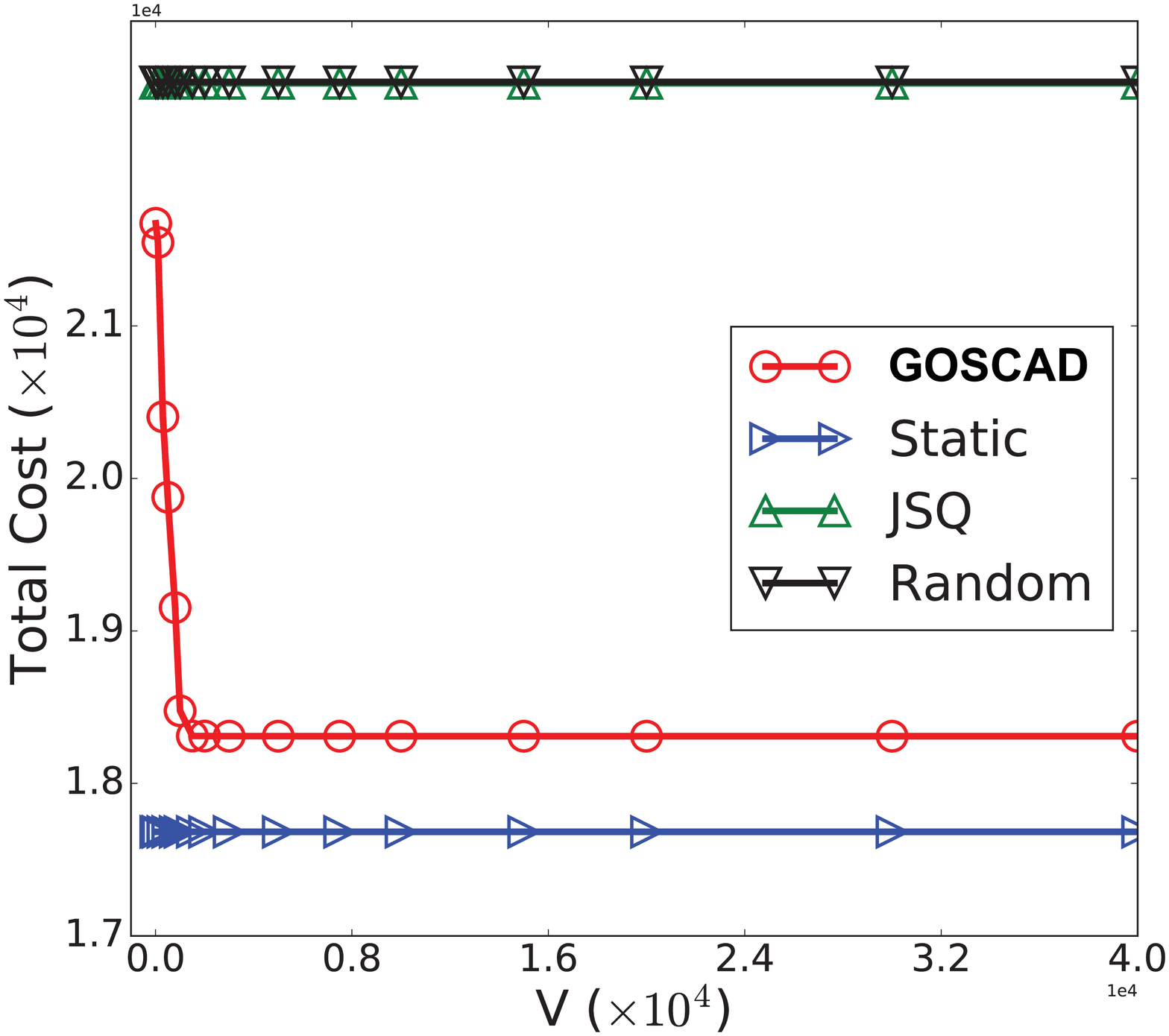}
 }
 \subfigure[Fat-tree topology] {
 \includegraphics[width=0.44\columnwidth]{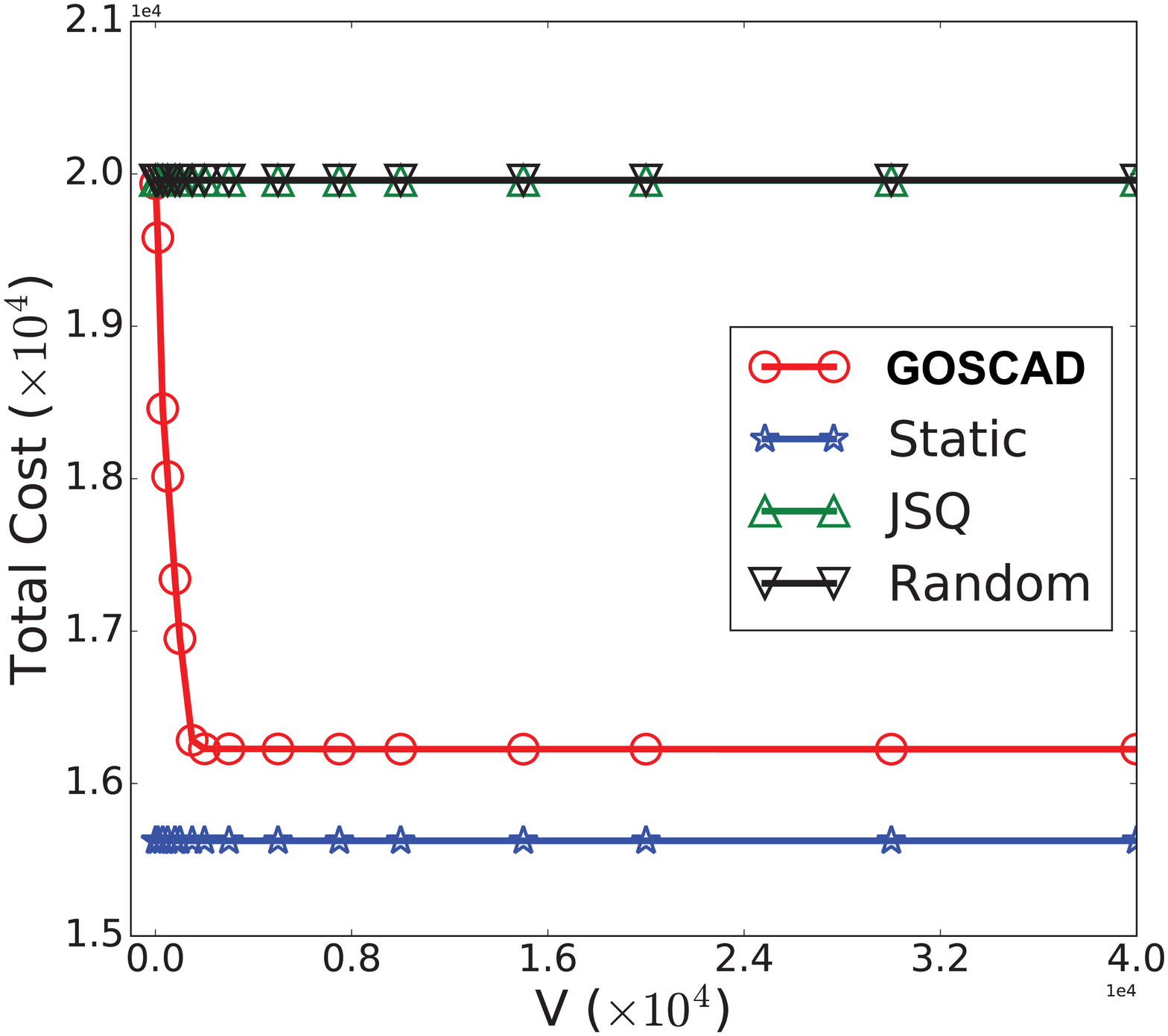}
 }
 \subfigure[Jellyfish topology] {
 \includegraphics[width=0.44\columnwidth]{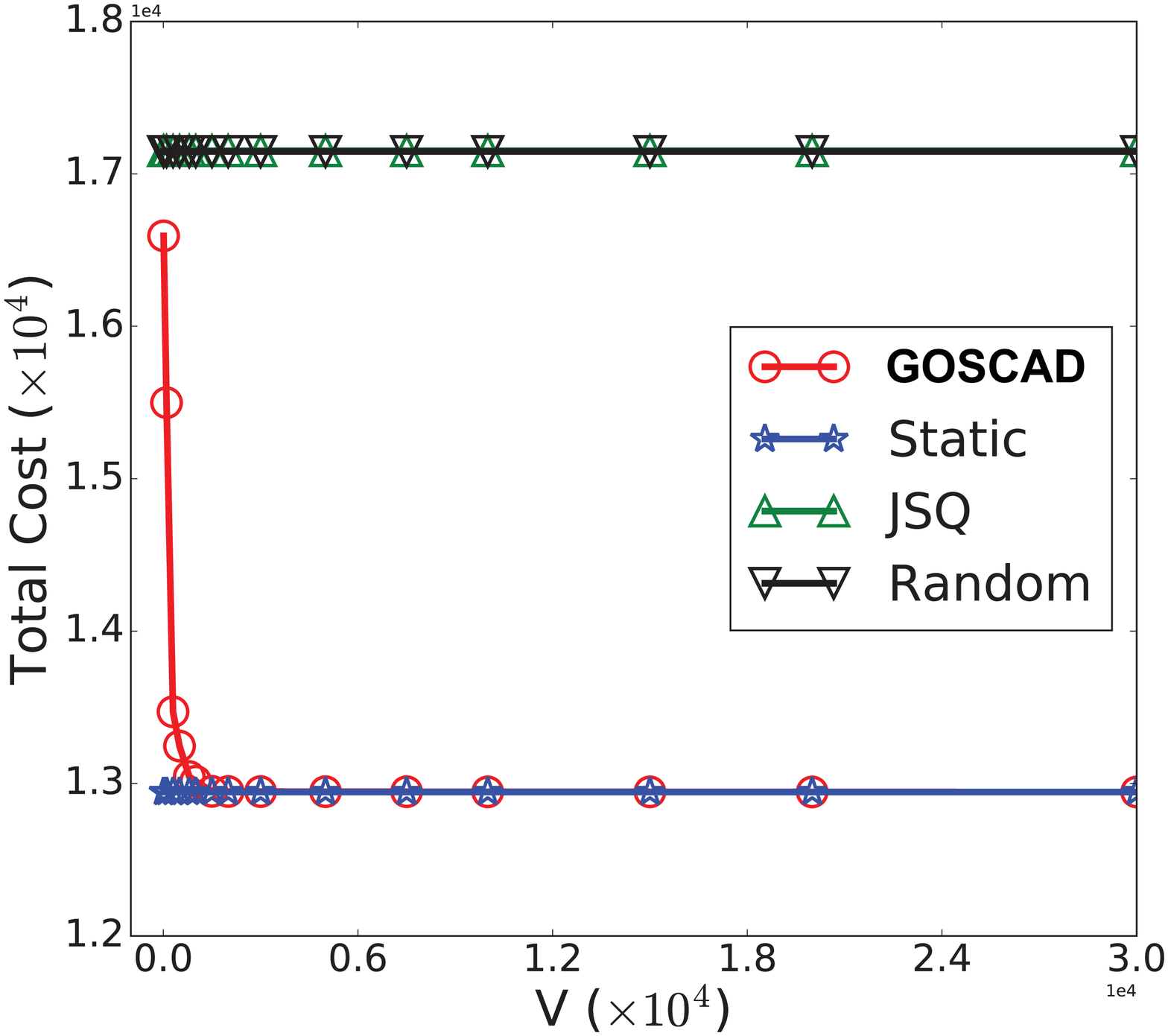}
 }
 \subfigure[F10 topology] {
 \includegraphics[width=0.44\columnwidth]{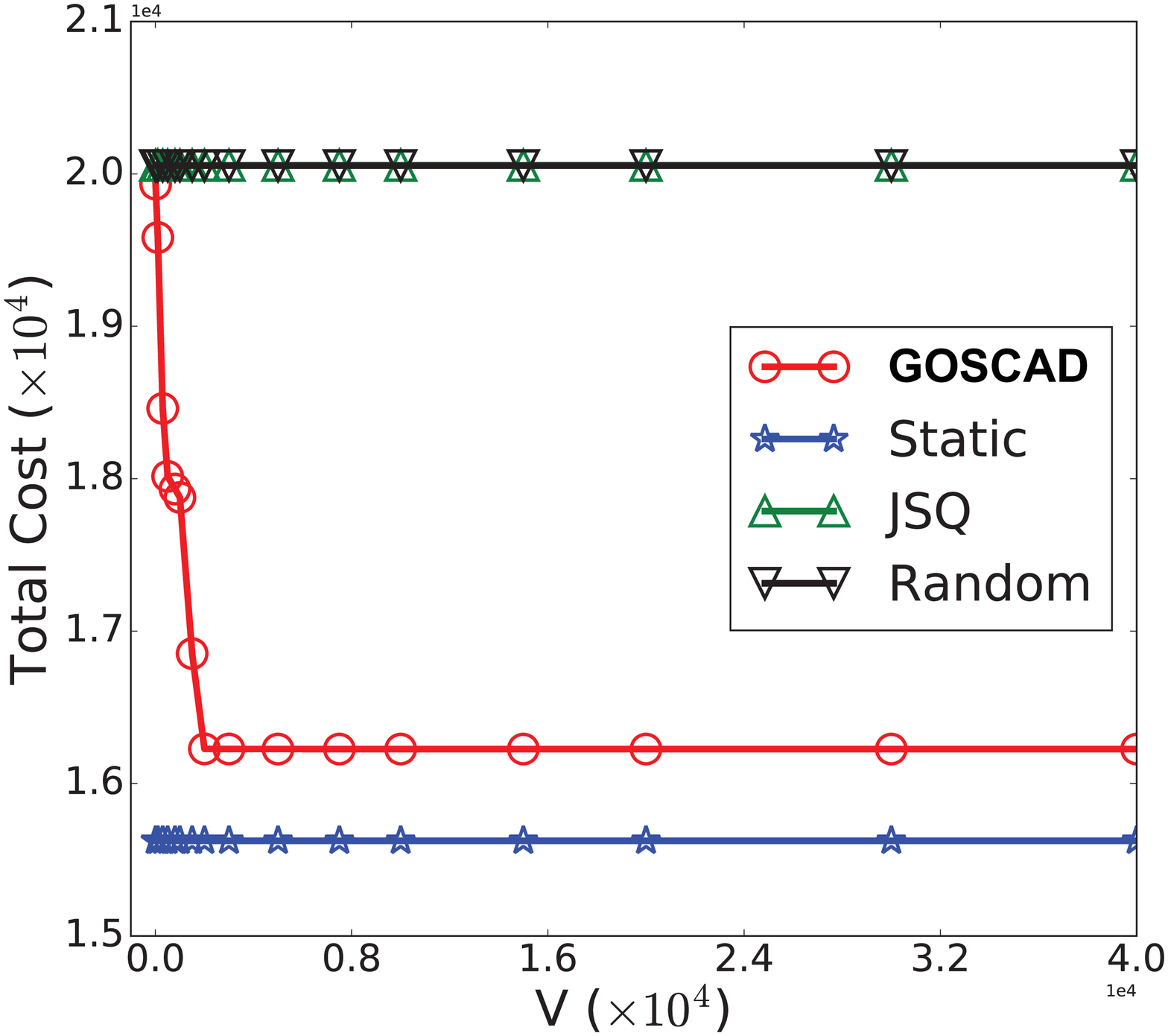}
 } 
 \caption{Comparison of communication costs among four scheduling schemes under different topologies. Note that \textit{GOSCAD} corresponds to the special case of \textit{POSCAD} without prediction.}
 \vspace{-2em}
 \label{cost_trace}
\end{figure}

\subsection{Performance Comparison against Baseline Schemes}

\begin{figure}[!t]
\centering
 \subfigure[Canonical 3-Tiered topology] {
 \includegraphics[width=0.44\columnwidth]{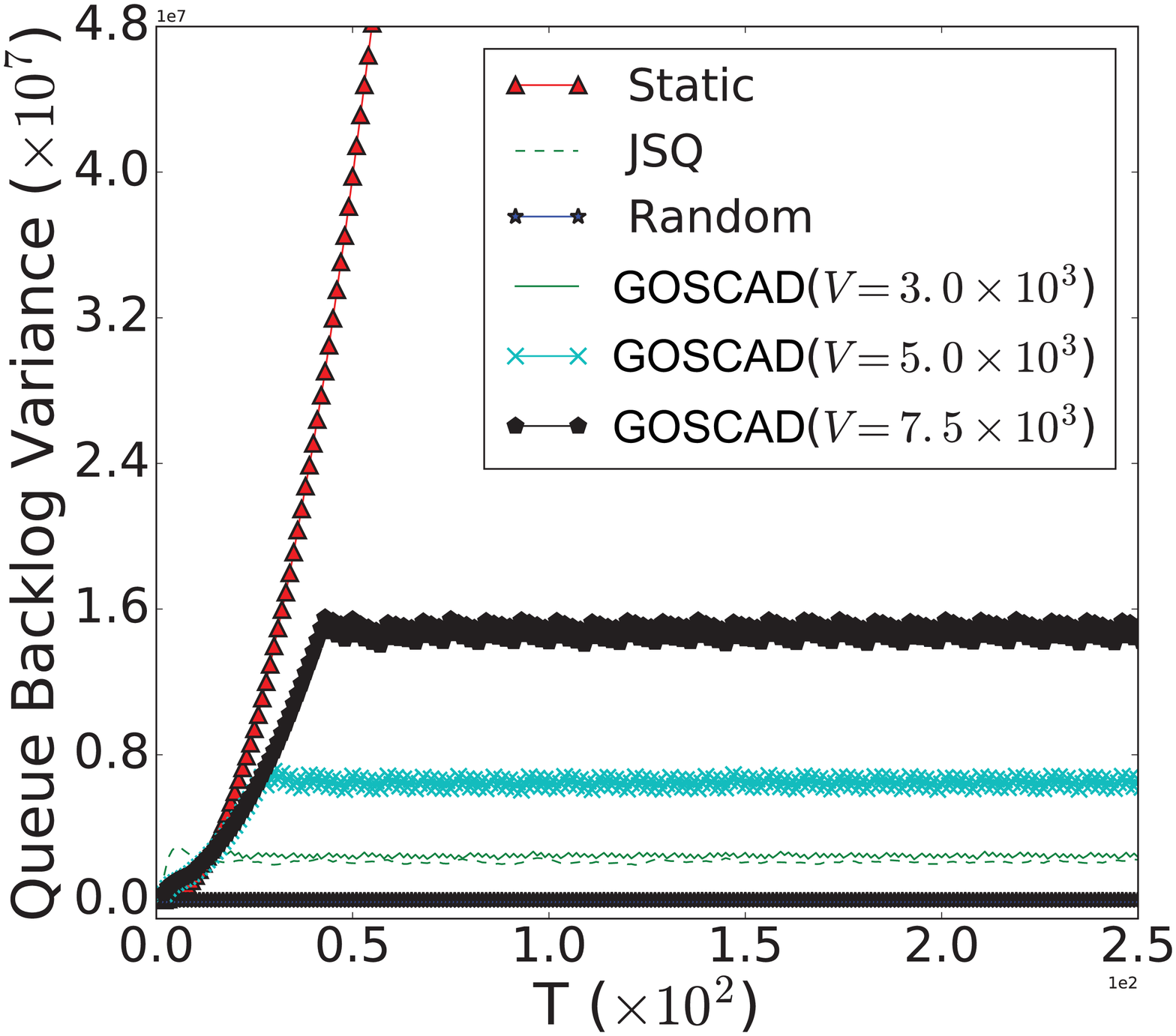}
 }
 \subfigure[Fat-tree topology] {
 \includegraphics[width=0.44\columnwidth]{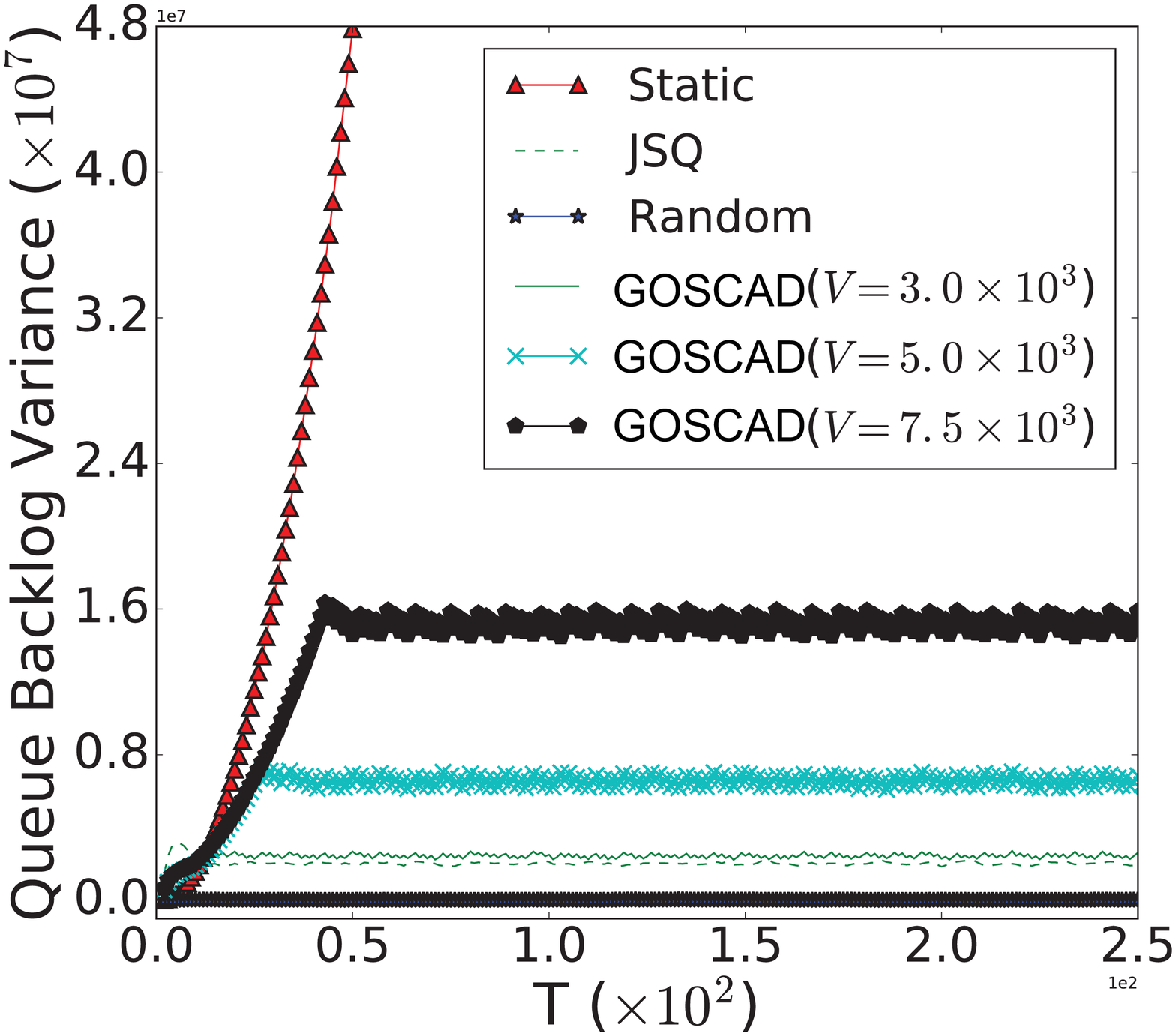}
}
\subfigure[Jellyfish topology] {
 \includegraphics[width=0.44\columnwidth]{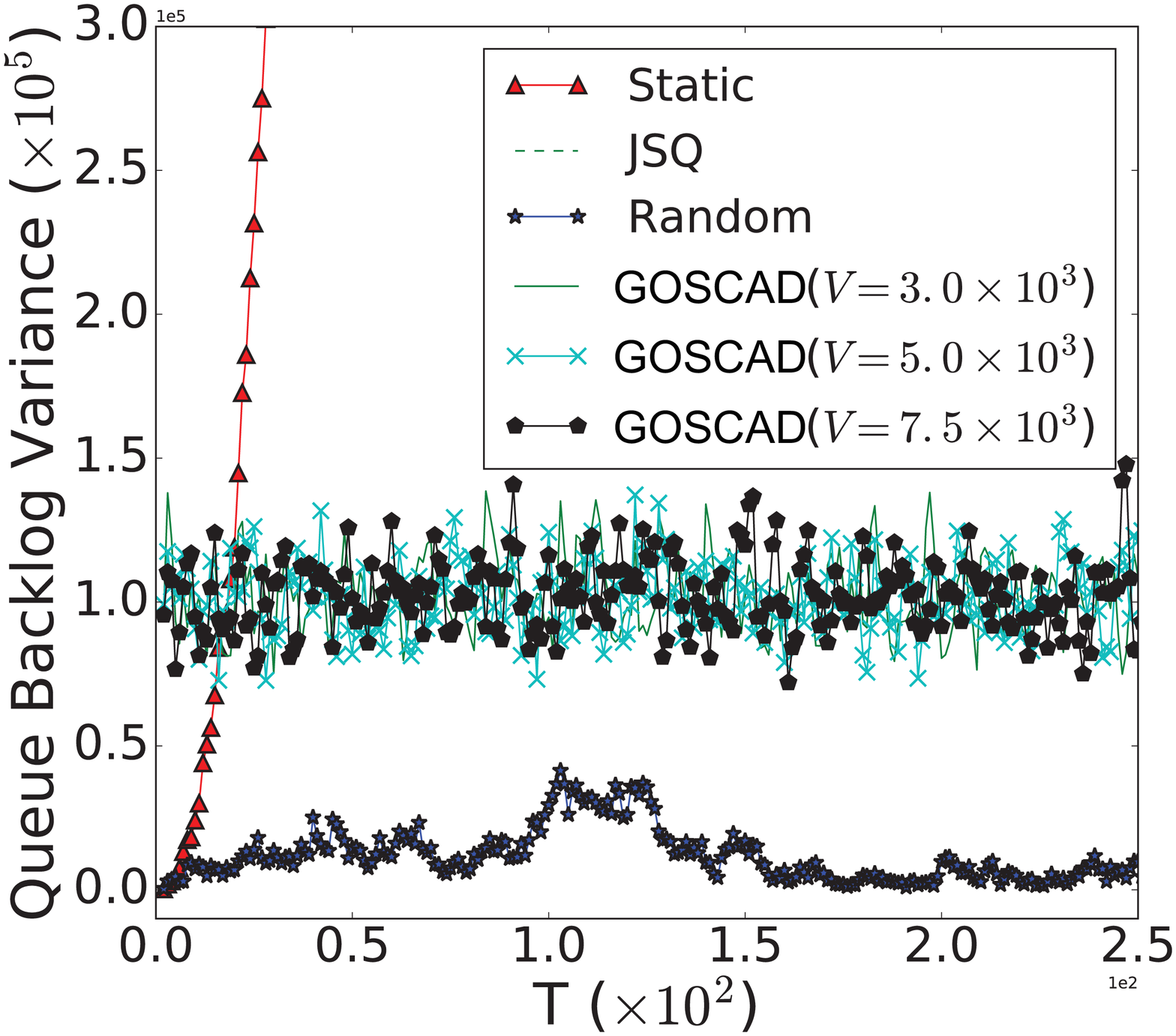}
 }
 \subfigure[F10 topology] {
 \includegraphics[width=0.44\columnwidth]{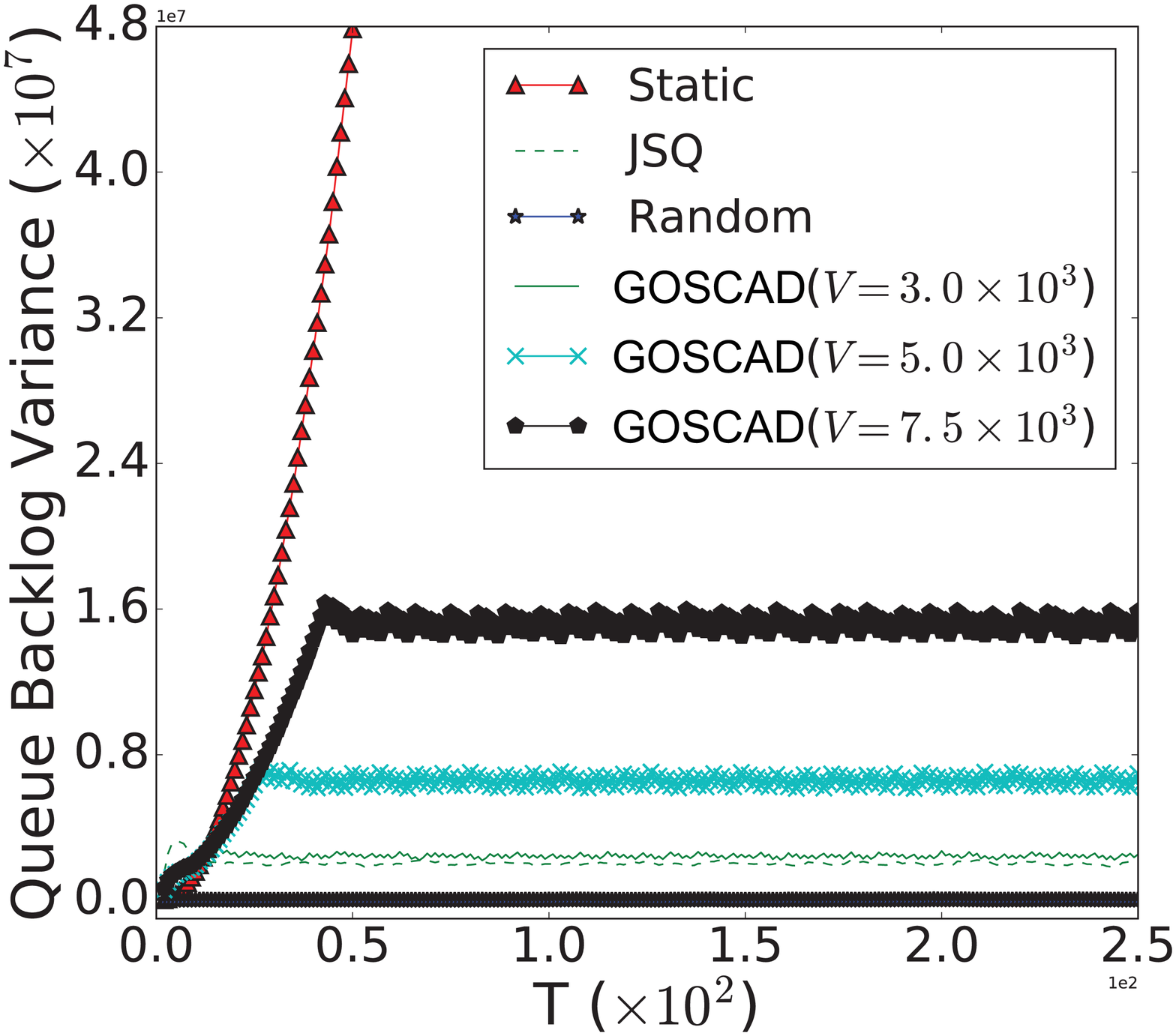}
}
 \caption{Variances of queue backlog size comparison incurred by four scheduling schemes under different topologies. Note that \textit{GOSCAD} corresponds to the special case of \textit{POSCAD} without prediction.}
 \vspace{-1.5em}
 \label{var_trace}
\end{figure}

\begin{figure}[!t]
\centering
 \subfigure[Canonical 3-Tiered topology] {
 \includegraphics[width=0.45\columnwidth]{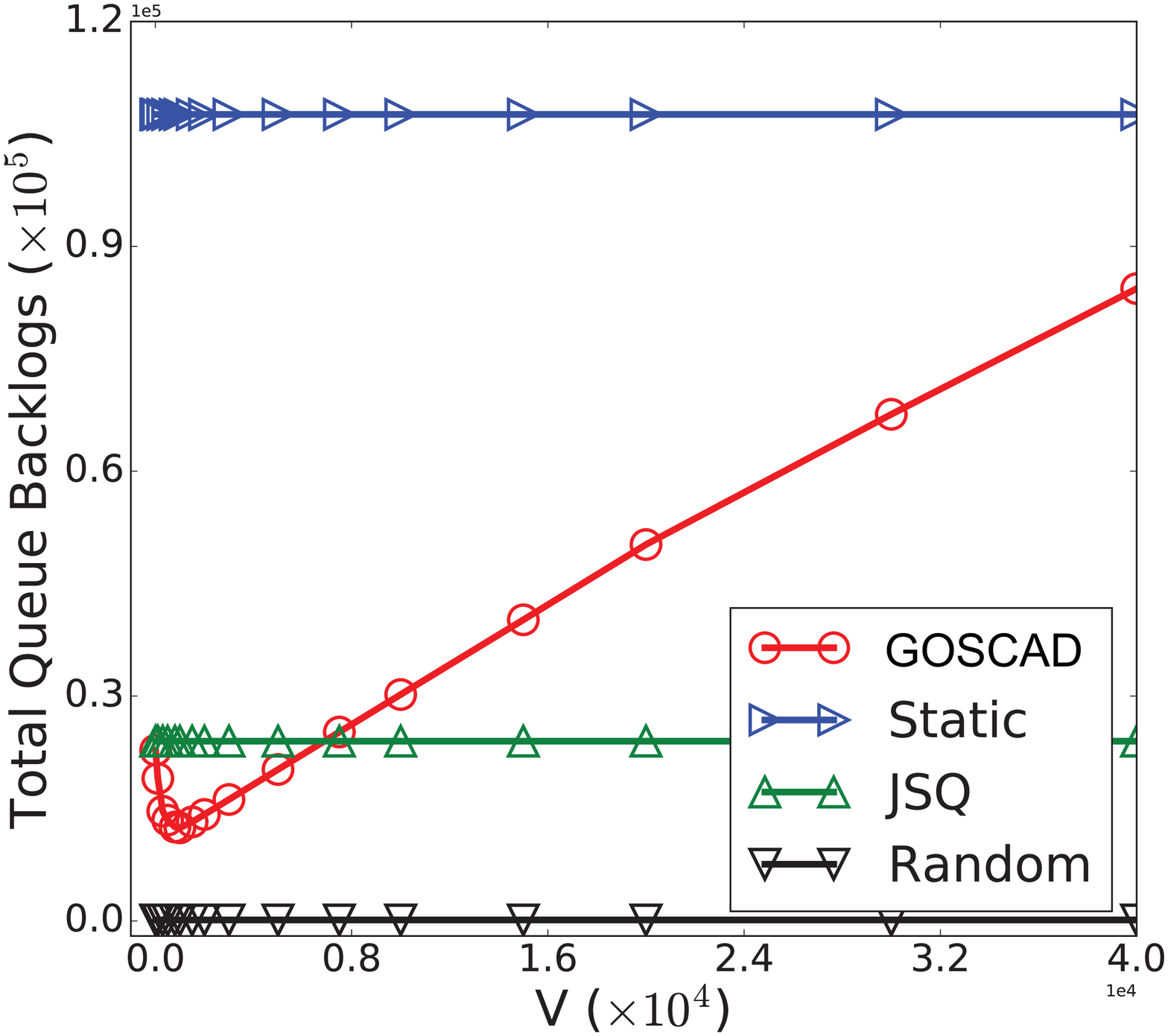}
 }
 \subfigure[Fat-tree topology] {
 \includegraphics[width=0.45\columnwidth]{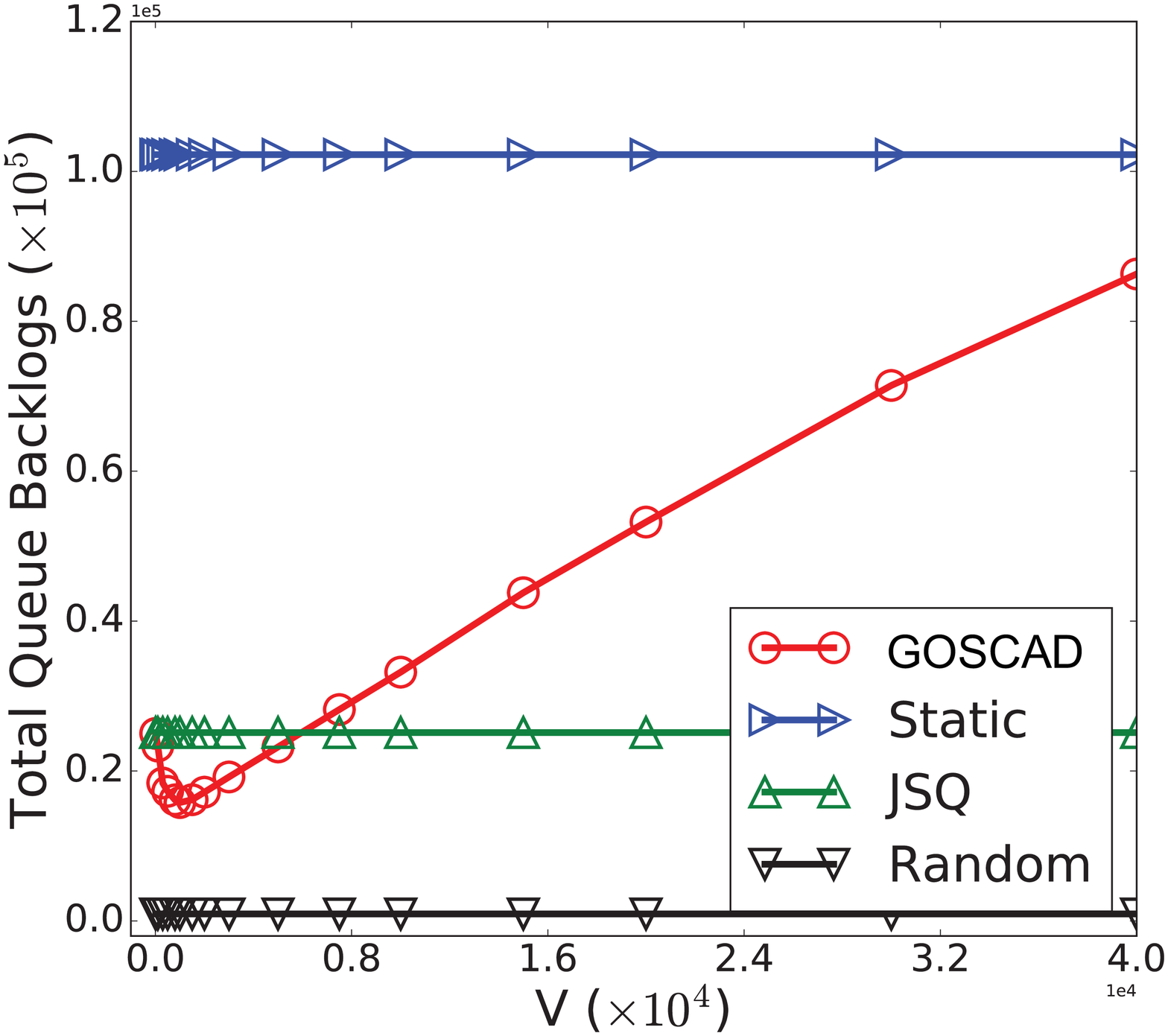}
 }
 \subfigure[Jellyfish topology] {
 \includegraphics[width=0.45\columnwidth]{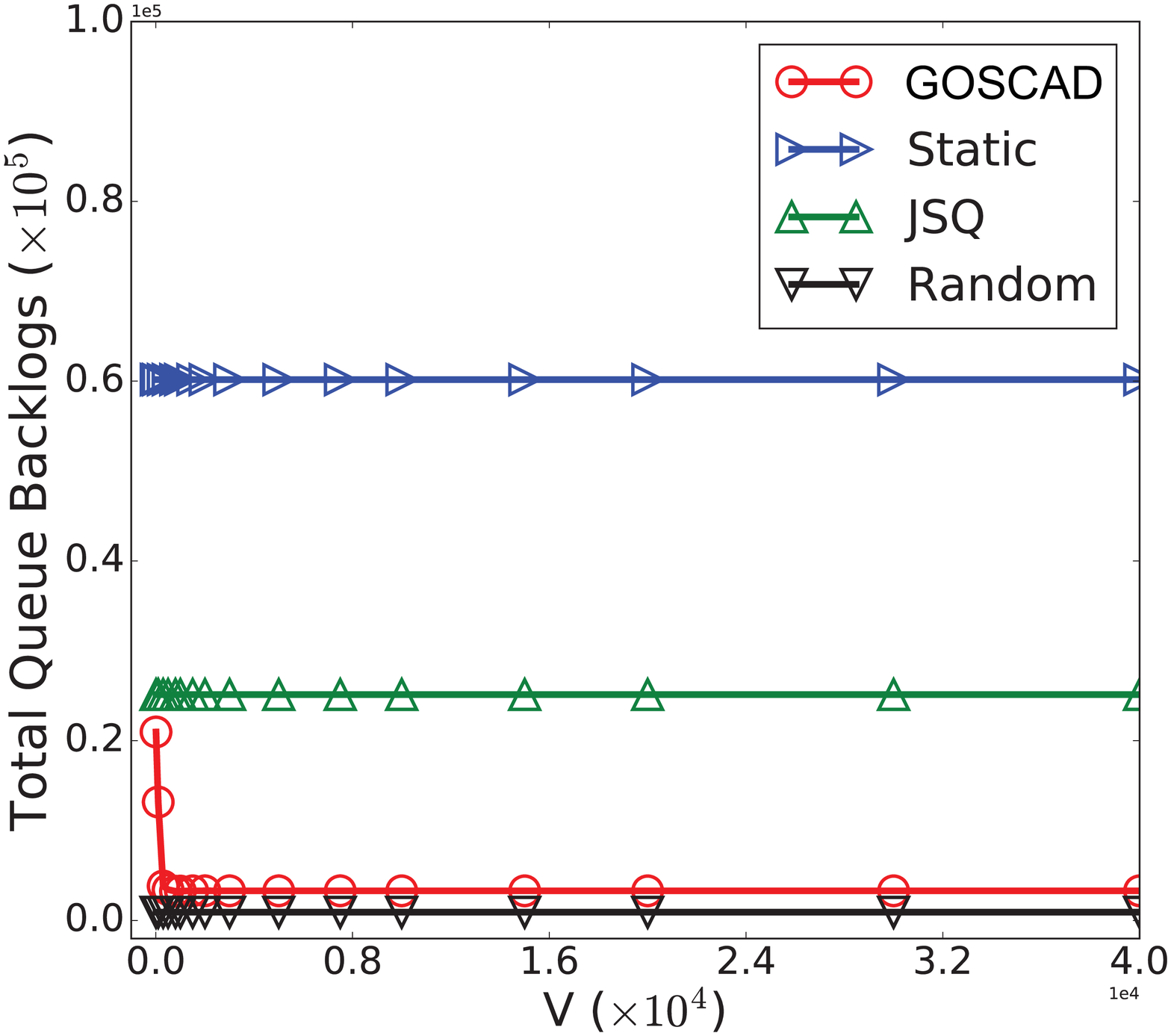}
 }
 \subfigure[F10 topology] {
 \includegraphics[width=0.45\columnwidth]{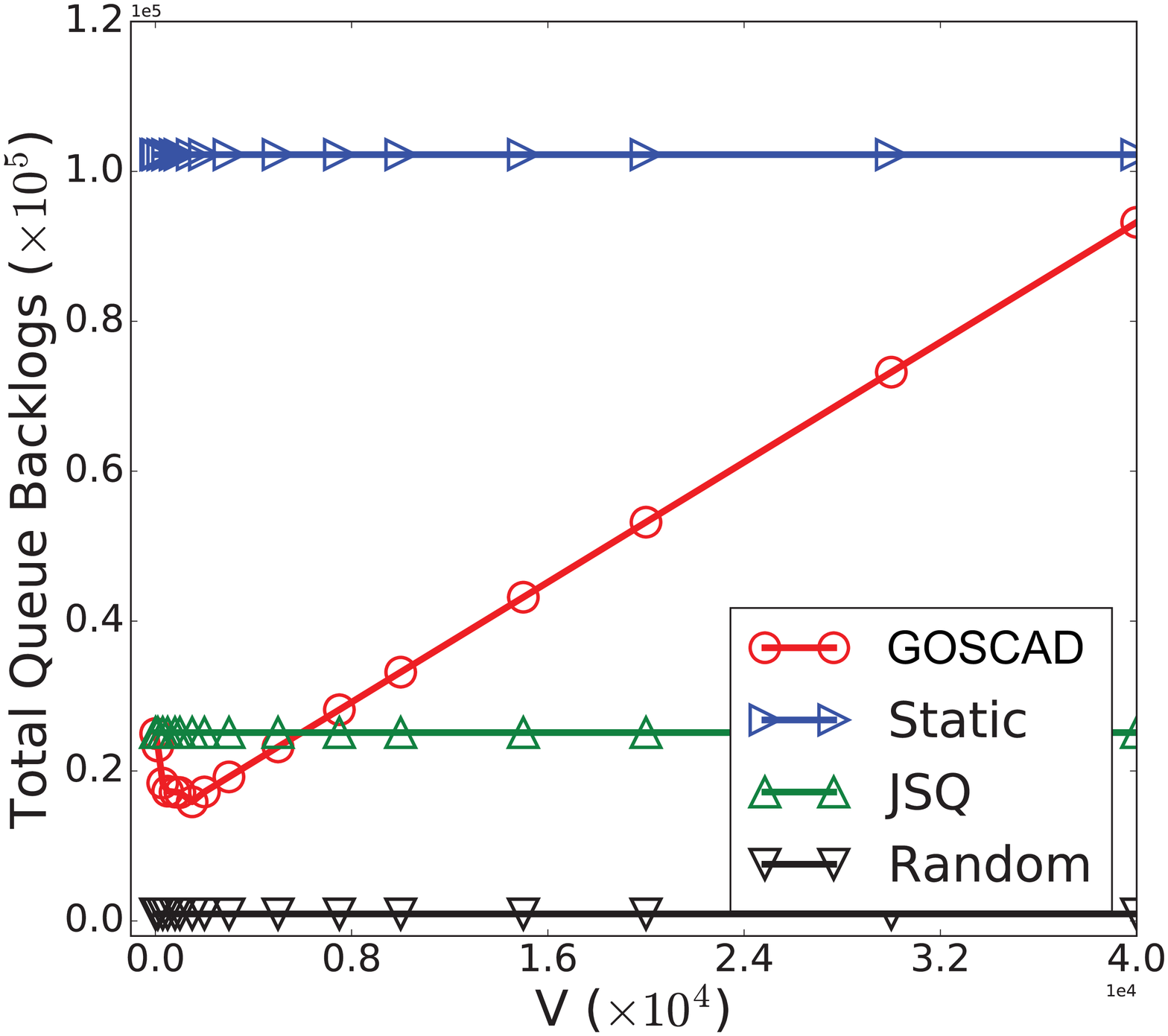}
 }
 \caption{Total queue backlog size comparison among four scheduling schemes under different topologies. Note that \textit{GOSCAD} corresponds to the special case of \textit{POSCAD} without prediction.}
 \vspace{-1.9em}
 \label{qlen_trace}
\end{figure}

\begin{figure}[!t]
\centering
 \subfigure[Poisson] {
 \includegraphics[width=0.44\columnwidth]{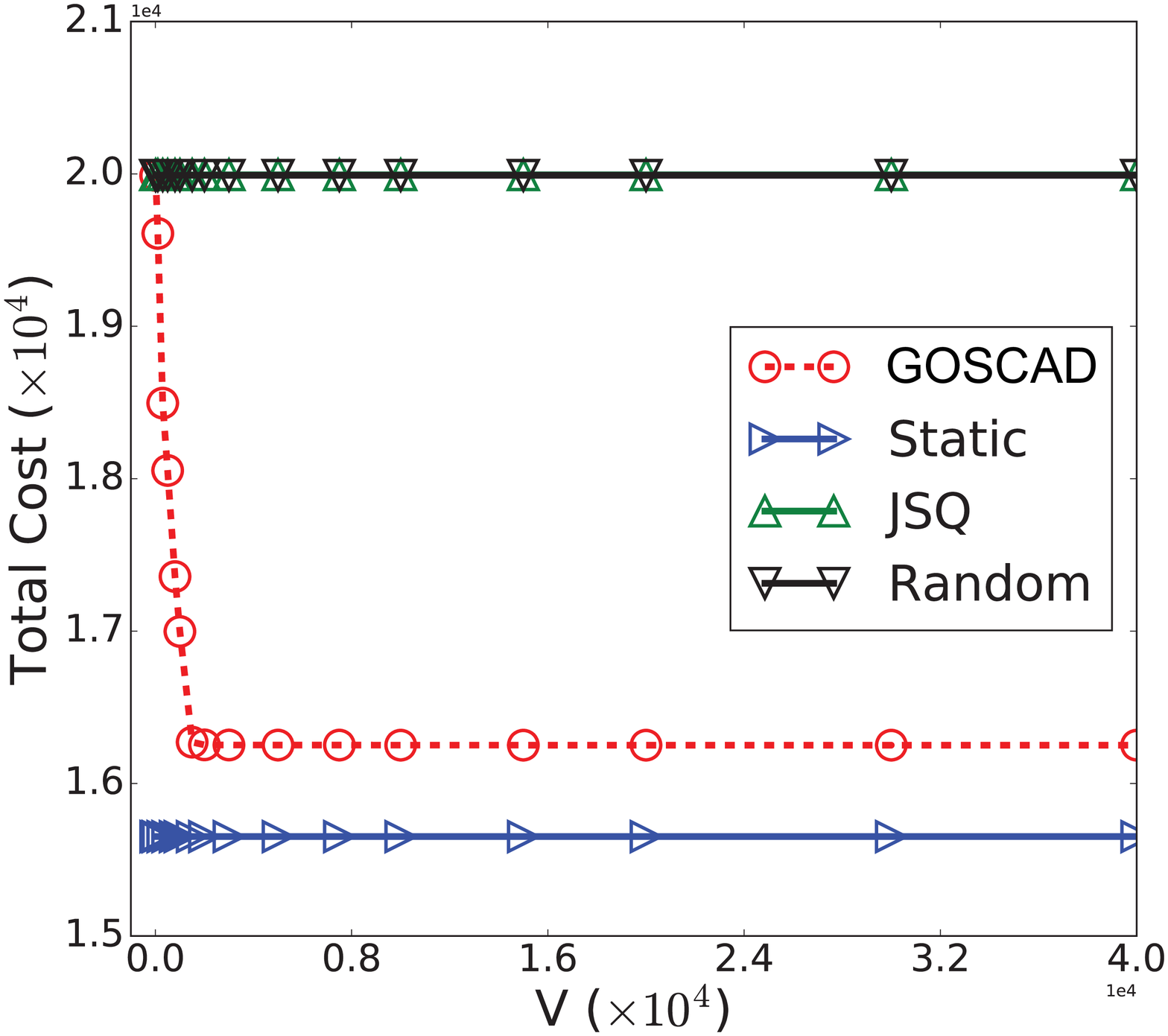}
 }
 \subfigure[Pareto] {
 \includegraphics[width=0.44\columnwidth]{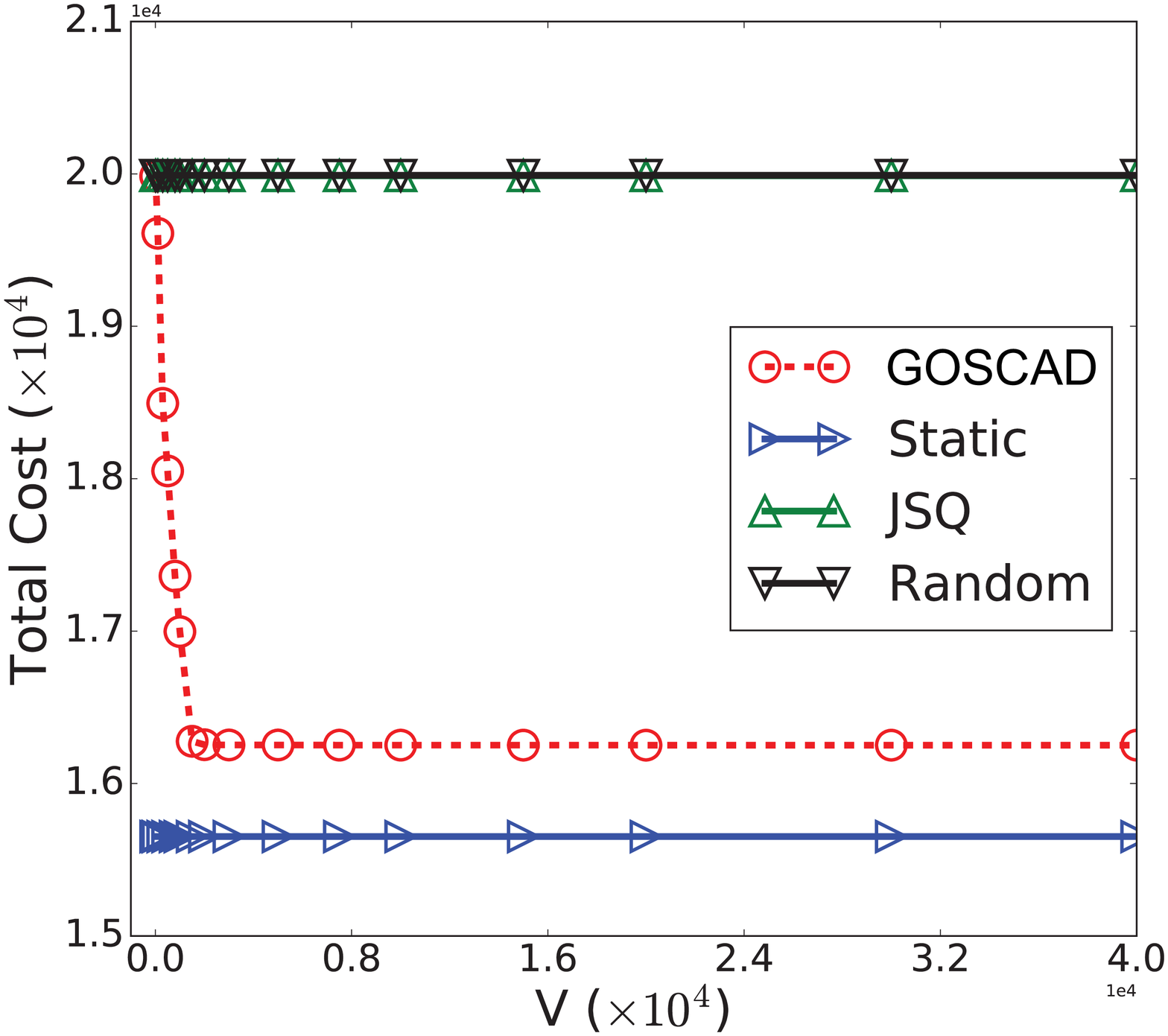}
 }
 \vspace{-0.2em}
 \subfigure[Poisson] {
 \includegraphics[width=0.44\columnwidth]{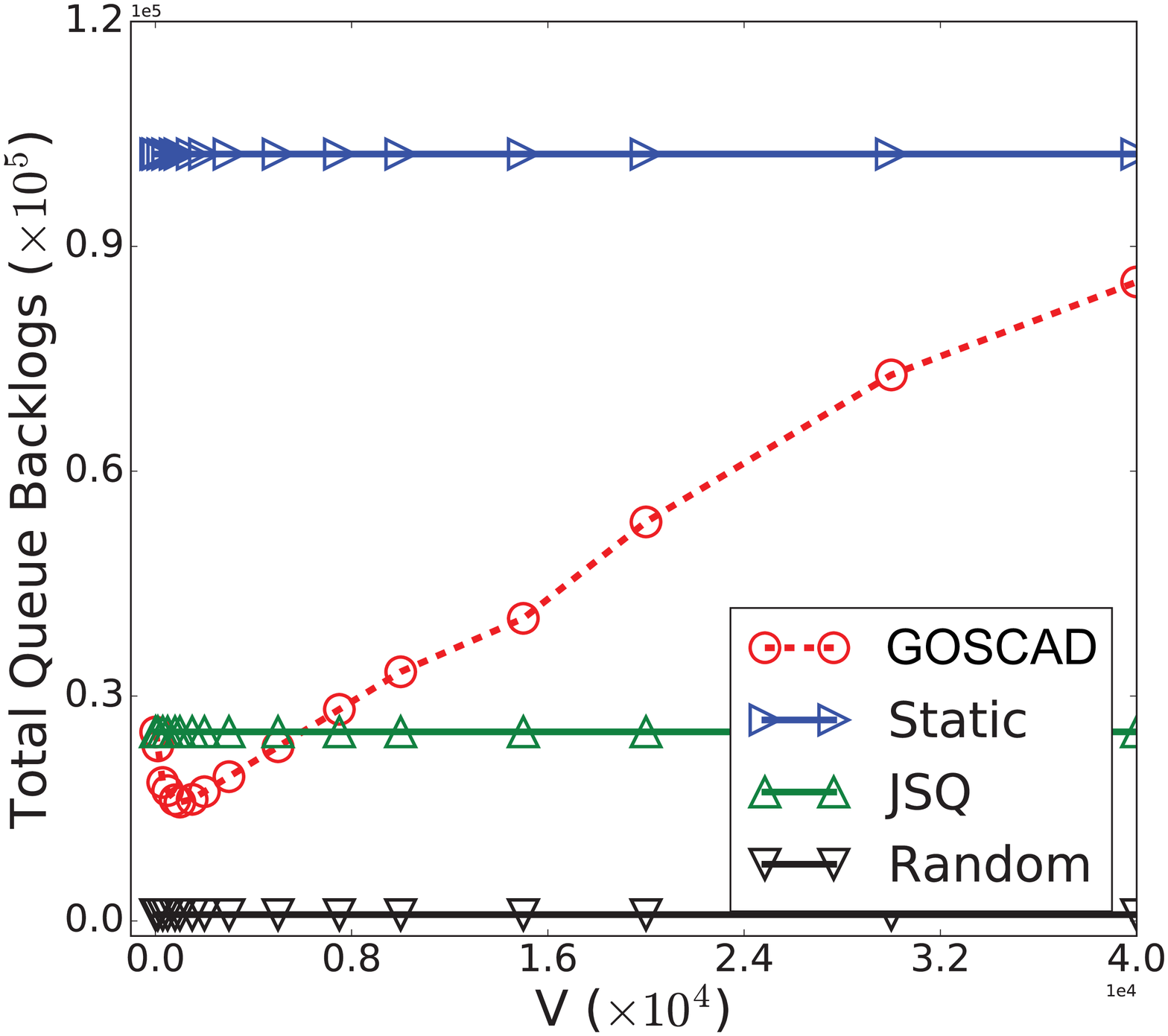}
 }
 \subfigure[Pareto] {
 \includegraphics[width=0.44\columnwidth]{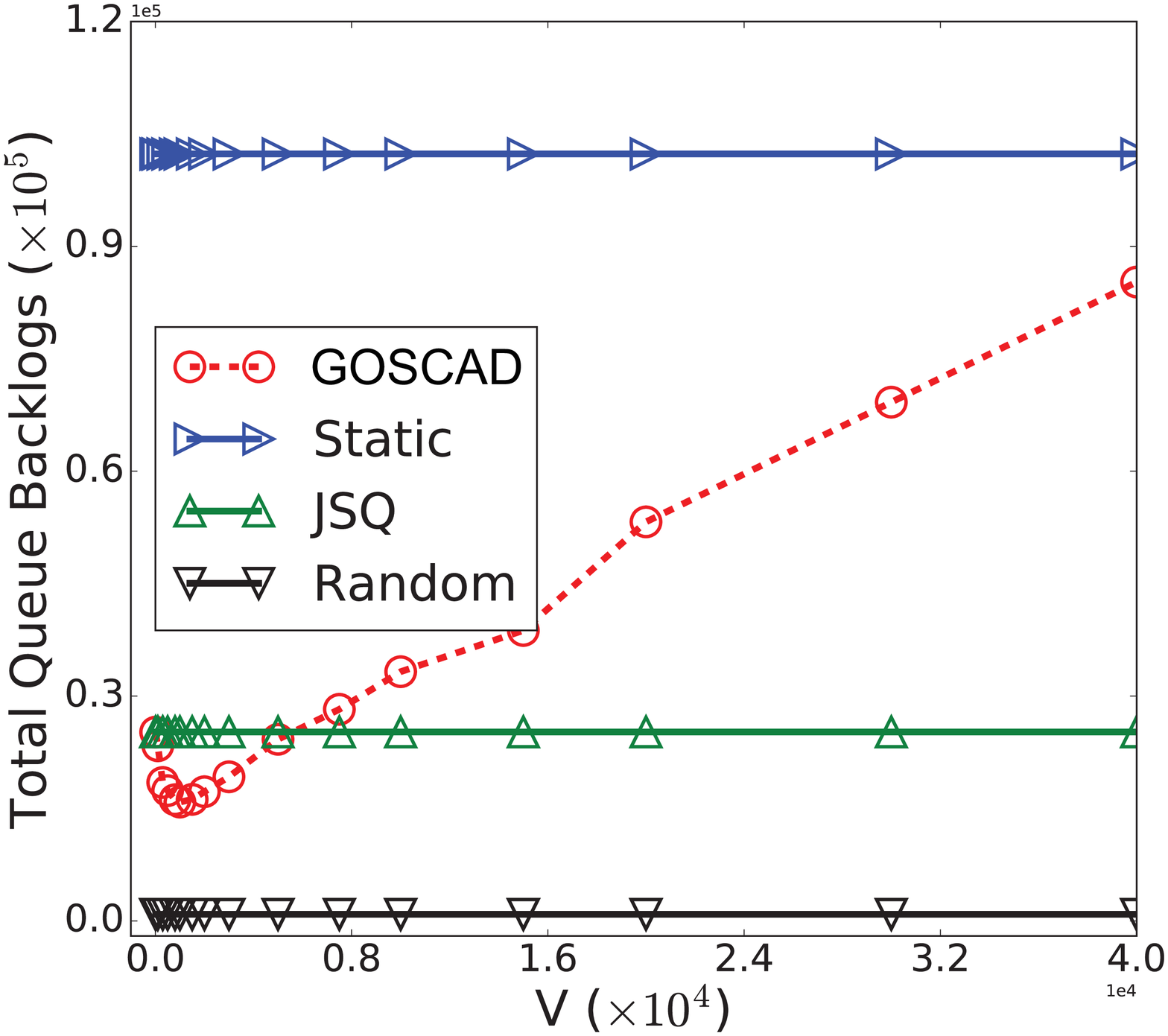}
 }
 \caption{Performance comparison among different schemes under Fat-tree topology, with Poisson and Pareto request arrivals, respectively. Note that \textit{GOSCAD} corresponds to the special case of \textit{POSCAD} without prediction.}
 \vspace{-1.8em}
 \label{pois-pareto}
\end{figure}

In Figures \ref{cost_trace} - \ref{pois-pareto}, we compare the performance of \textit{GOSCAD} (\textit{POSCAD} without prediction) against three baseline schemes: \textit{Static}, \textit{Random}, and \textit{JSQ} (Join-the-Shortest-Queue). Particularly, in \textit{Static} scheme, each switch $i$ chooses the controller $j$ with the minimum communication costs. 
In \textit{Random} scheme, each switch is scheduled to pick up a controller uniformly at random during each time slot. 
In \textit{JSQ} scheme, each switch picks the controller with the shortest queue backlog. 
Note that the baseline schemes do not conduct control devolution. To make the comparison fair, we set the unit computational cost $P = 10^{28}$ for all switches. 
This makes the costs of local computation prohibitively high; as a result, under \textit{GOSCAD}, switches would keep uploading new requests to associated controllers. That being said, \textit{GOSCAD} degenerates into a dynamic switch-controller association scheme. Note that such settings also emulate the scenarios in which control devolution is not supported.

From Figure \ref{cost_trace}, we see that \textit{Static} achieves the minimum total costs, since it greedily associates switches to controllers with minimum communication costs. On the other hand, both \textit{Random} and \textit{JSQ} incur much higher communication costs. 
The reason is that their decisions make no use of system dynamics about communication costs. Compared to baseline schemes, \textit{GOSCAD} cuts down the communication costs as the value of $V$ increases. 
Eventually, its induced communication costs stays $5.7\%$ above the curve of \textit{Static}. 
In fact, such a gap is the price taken to maintain queue stability. Note that when the value of $V$ increases, \textit{GOSCAD}'s behavior becomes increasingly similar to \textit{Static}, which focuses more on the reduction in total system costs. The difference emerges when some controllers' queue backlog sizes exceed some threshold (about twice the value of $V$ in our simulations).
In such cases, \textit{Static} would continue pursuing minimum communication costs and send requests to controllers which are heavily loaded but with the lowest communication costs. Consequently, such controllers will become overloaded and even lead to system breakdown. Under \textit{GOSCAD}, however, some switches would rather send requests to controllers with higher costs but a smaller queue backlog size, to ensure queue stability. Such illustration is supported by the results from Figure \ref{var_trace}. From the results, we see that the larger the value of parameter $V$, the greater the variance of queue backlog sizes among different controllers. 
In the extreme case (\textit{Static}), the variance grows remarkably. Unlike \textit{Static}, \textit{GOSCAD} constantly ensures queue stability. Besides, we also notice that such a gap is much smaller under Jellyfish topology, because therein the communication costs are more even to each switch, resulting in more balanced queue backlogs in the system.

\textbf{Insight}: As the value of $V$ increases, \textit{GOSCAD} achieves a tunable trade-off between system cost reduction and queue stability. The gap between \textit{GOSCAD} and the minimum costs is the price which has to be taken to maintain queue stability.

Figure \ref{qlen_trace} shows the induced total queue backlog sizes under GOSCAD and the baseline schemes. We see that under different topologies, both \textit{Random} and \textit{JSQ} maintain the total queue backlog sizes at a low level, which is consistent with the result in Figure \ref{var_trace}. In contrast, \textit{Static} induces significantly higher queue backlog sizes, due to its greediness to send requests to controllers with lowest costs, leading to imbalanced queue backlogs. Intuitively, the more balanced the queue backlogs across controllers, the more controller resources are utilized, and hence a smaller total queue backlog size.

As for \textit{GOSCAD}, under deterministic topologies (3-Tiered, Fat-tree, and F10), we observe a reduction in the backlog size at the very beginning, then a linear increase after reaching a valley at around $10^3$. 
The explanation is as follows. When the value of $V$ is small, switches prefer controllers with shorter queues.\footnote{
		Note that \textit{JSQ} is a special case of \textit{GOSCAD} with $V = 0$.
	} 
	Recall that a switch's scheduling decision is independent of the others'. 
	This will lead to requests being intensively forwarded to few controllers. 
	Then controllers close to hot spots are more likely to be heavily loaded. As the value of $V$ becomes larger, some switches would reach a tipping point and choose other controllers instead. As a result, this would make controllers' backlogs more balanced, and hence smaller total queue backlog sizes. When the value of $V$ continues to grow, switches turn to forward requests with the aim to minimize communication costs. As a result, requests will constantly accumulate on some particular controllers, leading to queue backlog imbalance across controllers and hence the growth in the total queue backlog size.

\vspace{-0.3em}

When \textit{GOSCAD} is applied in Jellyfish, its curve is very different from other three topologies. We can see the significant reduction at the beginning and then it stays at a low level thereafter. We find that for Fat-tree, F10, and Canonical 3-Tiered, they have higher variance in the number of hops between switches and controllers; as a result, with the existence of hot spots and aim to minimize communication costs, more requests accumulate on some particular controllers, inducing increased queue backlog sizes. For Jellyfish, however, because incoming requests are spread more evenly among controllers, increasing the value of $V$ has no significant impact on the skewness of controller queue backlogs.

\textbf{Insight:} In practice, one can tune the value of parameter $V$ to be proportional to the ratio between the magnitude of queue backlog size and communication cost, to achieve a significant reduction in the total system costs while maintaining balanced queue backlogs among controllers. 

In addition, we also conduct simulations under two kinds of request arrival processes, \textit{i.e.}, \textit{Poisson} and \textit{Pareto} processes, which are widely adopted in traffic analysis. We only show the simulation results under Fat-tree topology, because the simulation results in the other three topologies are qualitatively similar. From Figure \ref{pois-pareto}, we find that all schemes perform qualitatively consistent under different arrival processes.

\subsection{Evaluation of POSCAD}

\subsubsection{Prediction Settings}
We vary switches' prediction window sizes by sampling them uniformly at random in $[0, 2 \times D]$, with mean $D$. The values of $D$ range from $0$ to $20$. 
Besides, considering that prediction errors are inevitable in practice, we evaluate POSCAD's performance with prediction errors at different levels. 
We use $e_t$ to denote the prediction deviation for time slot $t$, \textit{i.e.}, the difference between the number of predicted and actual arrivals. 
Particularly, $e_t = 0$ indicates that the prediction is perfect. 
When the prediction is imperfect, $e_t > 0$ implies that the future request arrivals are over-estimated; otherwise, they are under-estimated. We also assume $e_t$ to be \textit{i.i.d.} over  time slots, and generated by 1) sampling a value $x_t$ from some probability distribution with zero mean and then 2) obtaining $e_t$ by rounding $x_t$, \textit{i.e.}, $e_t \triangleq \text{Round}(x_t)$. 
The second step is due to that the probability distribution may be continuous but the number of requests must be an integer. Then we define prediction error rate $r$, \textit{i.e.}, the probability of mis-prediction, as $r \triangleq \text{Pr}\{e_t \neq 0\} = 
\text{Pr}\{x_t \notin (-0.5, 0.5)\}$. In our simulation, we sample $x_t$ from a \textit{normal distribution} with zero mean and variance $\sigma^2$. Accordingly, we have
\begin{equation}\label{def_r}
	r = 2[1-\Phi(0.5/\sigma)],
\end{equation}
where $\Phi(\cdot)$ is the CDF of \textit{standard normal distribution}. Based on (\ref{def_r}), the error rate $r$ can be adjusted by choosing an appropriate value of $\sigma$. We pick error rates from $0\%$ to $50\%$.

\subsubsection{Response Time Metric} 
As for evaluation metrics, note that the total queue backlog size in the non-prediction case $(D = 0)$ and the predictive case $(D > 0)$ are incomparable. The reason is that with predictive scheduling, the total queue backlog size, as defined in (\ref{def_total_q}), also includes future untreated requests. 
All such requests are still counted, although they have not actually arrived or even not been pre-served. The comparison is even more inappropriate with predictive scheduling in the presence of errors, where some future requests may not even exist due to over-estimation. Hence, instead of total queue backlog size, we evaluate \textit{POSCAD} in terms of average request response time, since \textit{POSCAD} is promising to reduce request response times by exploiting future information and pre-serving future requests with idle system resources. 
In our simulations, we define a request's response time as the number of time slots from its actual arrival to its eventual completion. If pre-served before its actual arrival, a request will be responded instantly and hence experience a near-zero response time. In our simulations, the average request response time is obtained over completed requests.

\subsubsection{Evaluation with Perfect Prediction}
To explore the benefits of predictive scheduling, we first consider the case where switches have perfect information about future request arrivals in the lookahead window, \textit{i.e.}, with error rate $r = 0$. Considering the similarity between curves under Fat-Tree and F10 topologies, we omit the results for F10 in the following.

\begin{figure}[!t]
    \begin{center}
    \subfigure[Window Size $D = 0$]{
        \includegraphics[scale=.16]{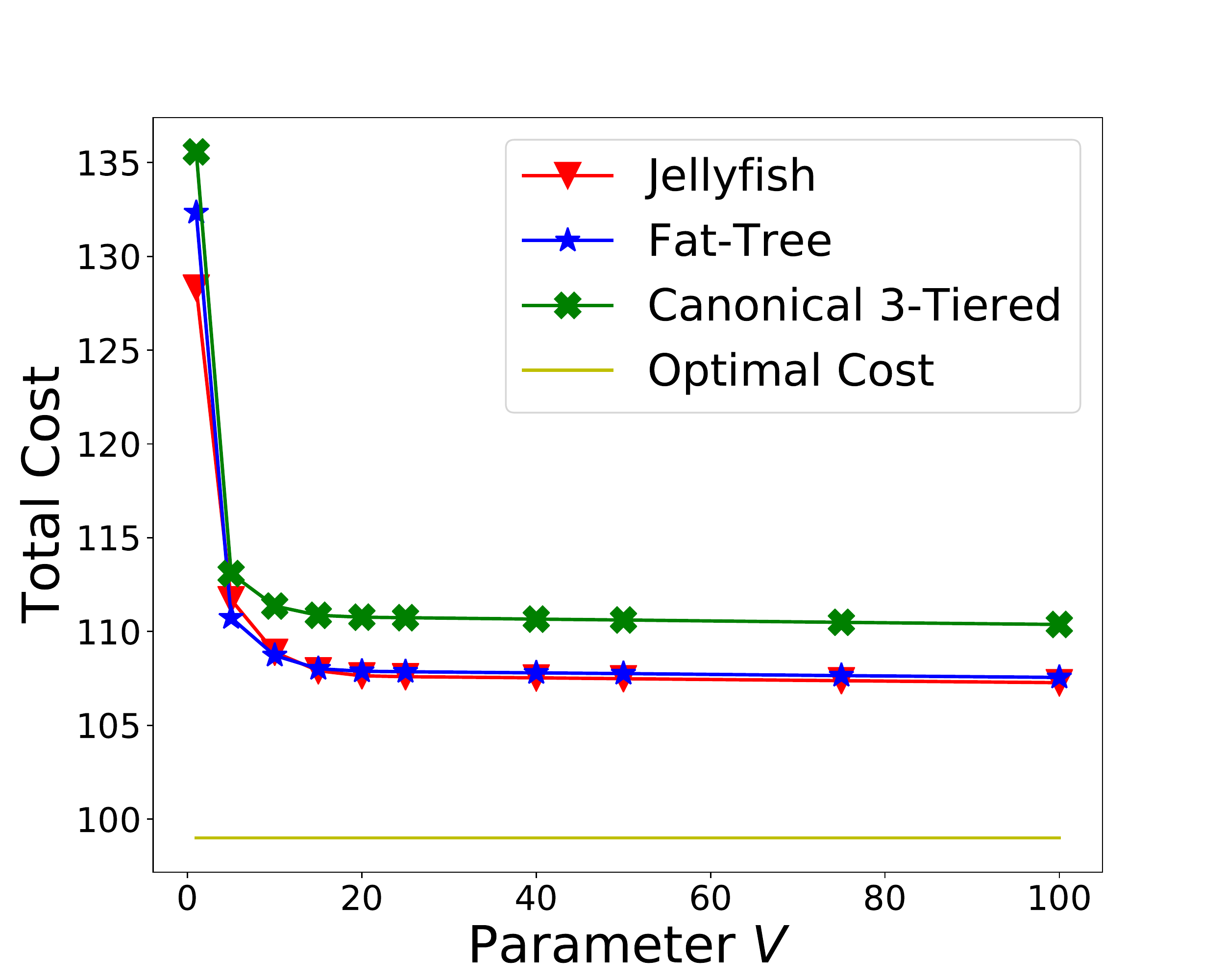}
    }
    \subfigure[Window Size $D = 2$]{
        \includegraphics[scale=.16]{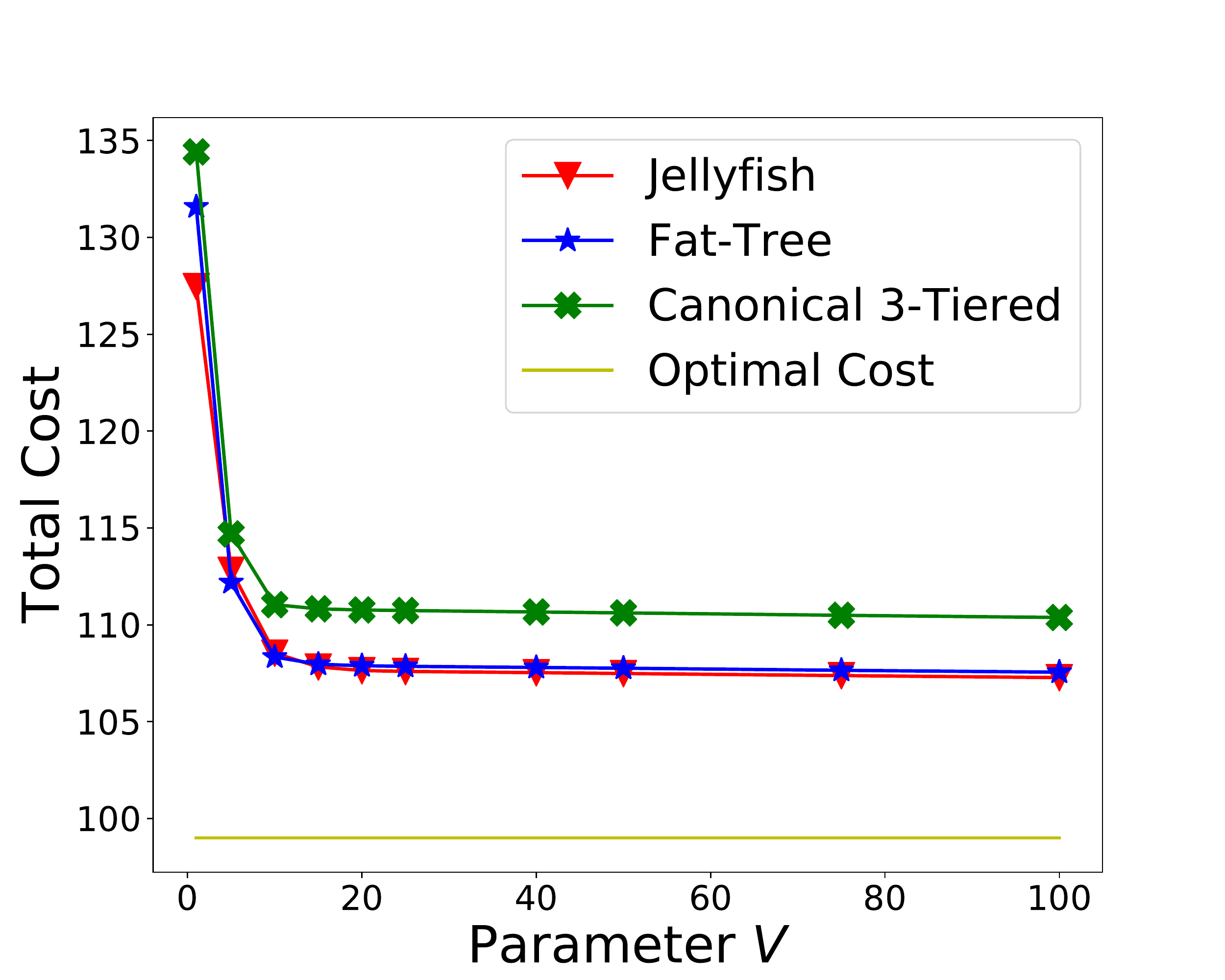}
    }
	\subfigure[Trace-driven]{
        \includegraphics[scale=.16]{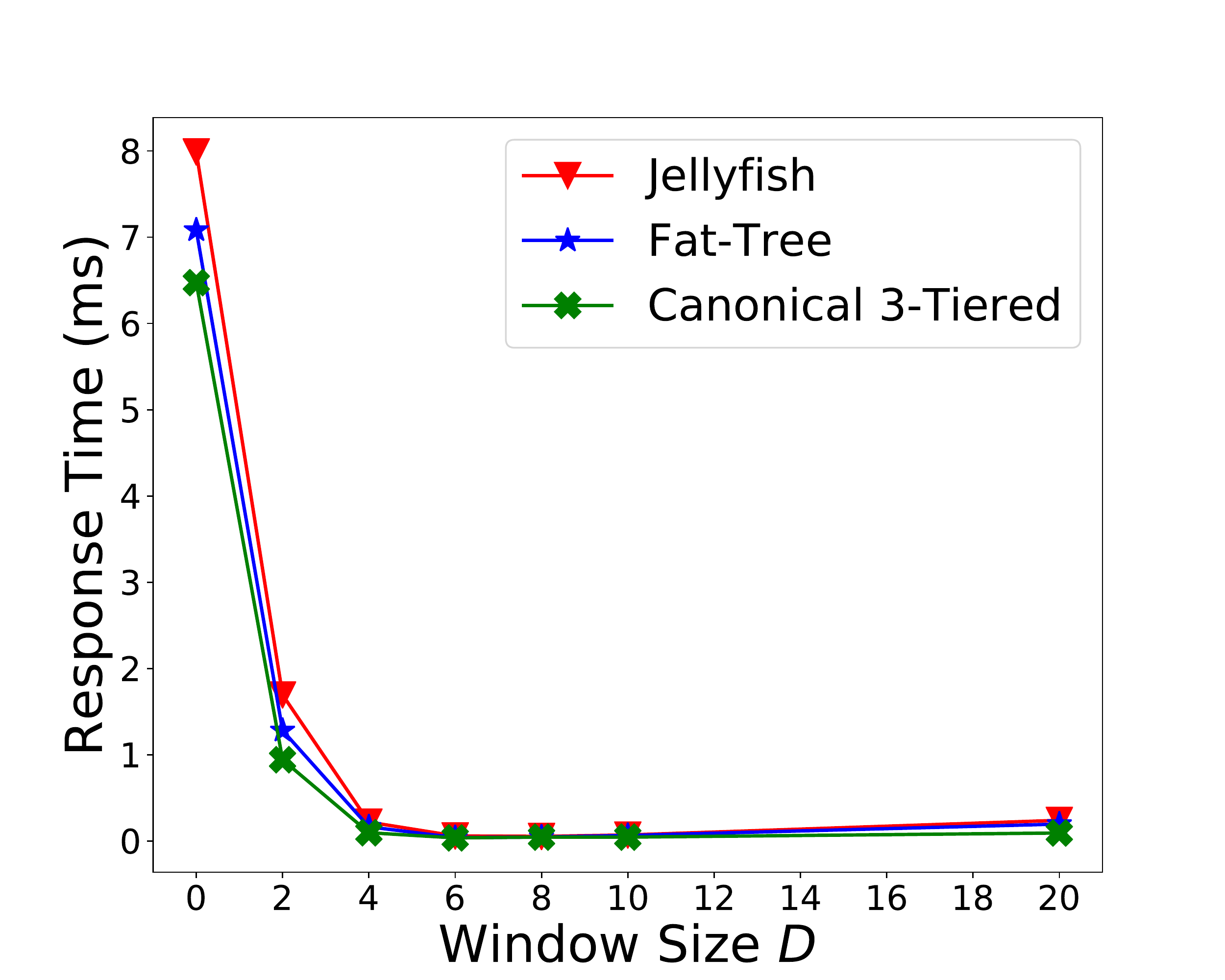}
    }
    \subfigure[Poisson]{
        \includegraphics[scale=.16]{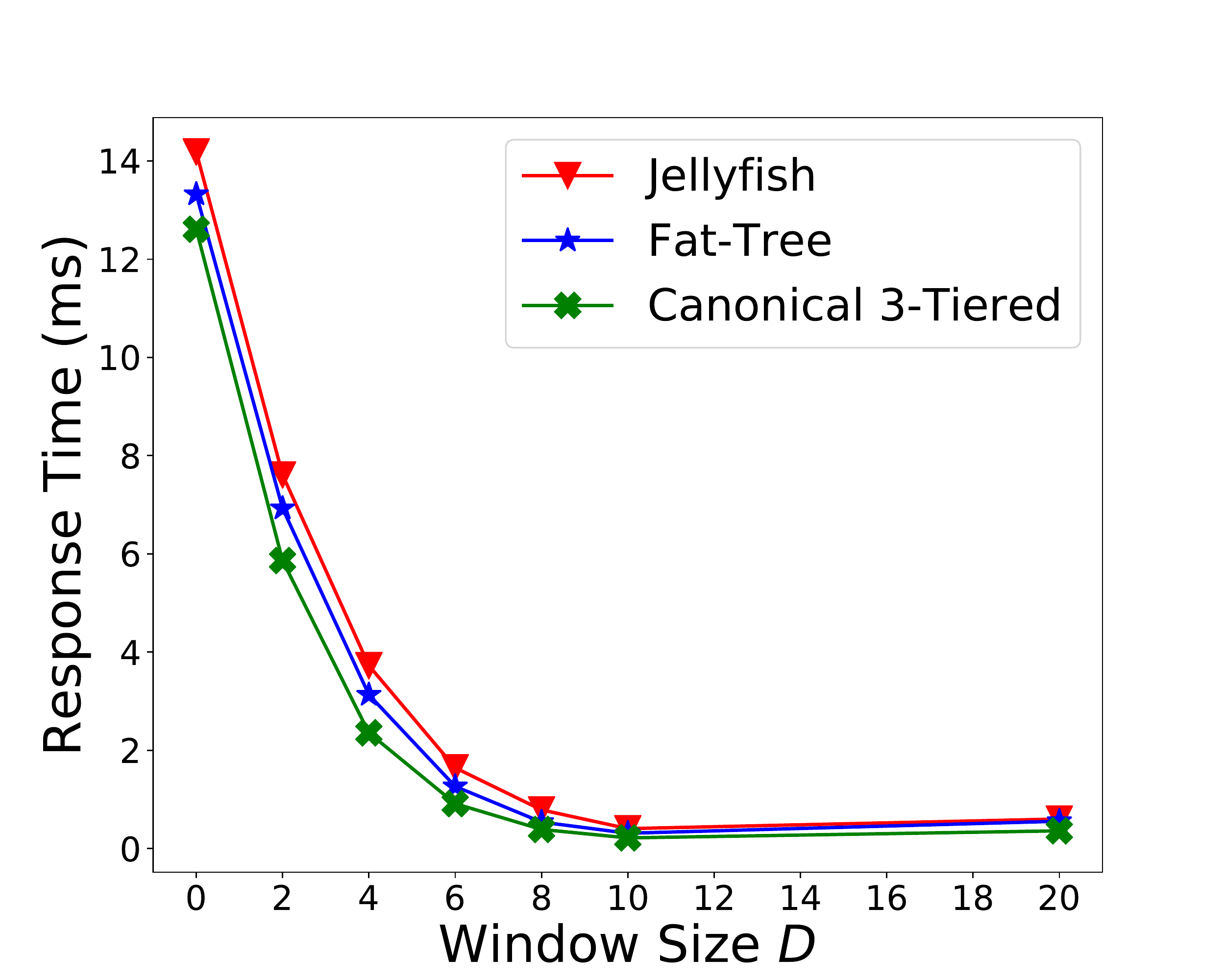}
    }
    \end{center}
    \caption{
		Performance under perfect prediction settings.
    }
    \label{perf_perfect_pred}
\end{figure}

\textbf{Time-averaged total costs vs. $V$}: 
We compare the total costs incurred by \textit{POSCAD} with different values of $V$ under different topologies in Figure \ref{perf_perfect_pred}. 
Particularly, we compare the results with prediction window sizes $D = 0$ (non-prediction) and $D = 2$ in Figures \ref{perf_perfect_pred} (a) and \ref{perf_perfect_pred} (b), respectively. 
We see a rapid reduction in total system costs when the value of $V$ increases from $1$ to $20$. As the value of $V$ continues, the reduction eventually diminishes with $10\%$ gap to the optimal total costs (denoted by the solid horizontal line). Comparing Figures \ref{perf_perfect_pred} (a) and \ref{perf_perfect_pred} (b), we see that the incurred total costs are almost the same even when the window size $D$ is increased. With such observations, we have the following insight.

\textbf{Insight}: As the value of parameter $V$ increases, \textit{POSCAD} reduces the time-averaged total costs and achieves the optimal system costs asymptotically under different topologies. Note that the gap between the converged costs and the optimal costs is the price paid for stabilizing queue backlogs in the system. In the meantime, compared to the non-prediction case, \textit{POSCAD} performs predictive scheduling without increasing the total costs in the system.

\textbf{Average response time vs. prediction window size}: 
Next, we focus on the impact of different prediction window sizes on the average request response time. Figures \ref{perf_perfect_pred} (c) and \ref{perf_perfect_pred} (d) present the results from simulations with trace-driven and Poisson arrival process, each with mean arrival rate $5.88$ requests per time slot, respectively. Figure \ref{perf_perfect_pred} (c) shows the curves under trace-driven settings. 
We see a sharp decrease of the average request response time from around $14$ms to less than $1$ms, as the window size rises from $0$ (non-prediction) to $6$ under different topologies. 
As the window size continues growing, the reduction stops and the average response time converges ($0.45$ms). The results are similar in Figure \ref{perf_perfect_pred} (d).

\begin{figure}[!t]
    \begin{center}
    \vspace{-2.3em}
    \hspace{-0.5em}
        \includegraphics[scale=.26]{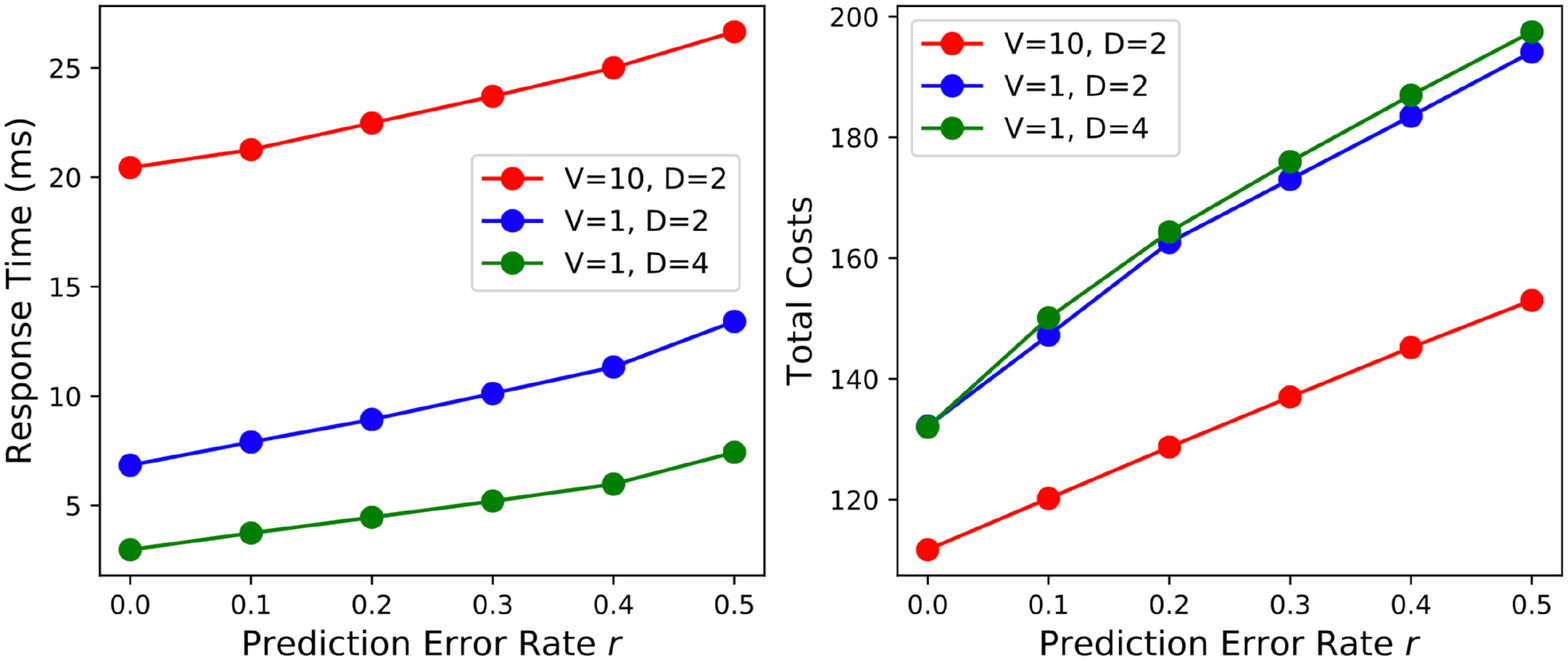}
	\vspace{-4em}
    \end{center}
    \caption{Average response time vs. prediction error rate under Fat Tree topology with different parameter $V$ and prediction window size $D$.}
    \label{fig_acc}
\end{figure}

\textbf{Insight:}
Figure \ref{perf_perfect_pred} shows the benefits of predictive scheduling. By exploiting predictive information to pre-serve requests, \textit{POSCAD} effectively reduces the average response time without incurring extra system costs. Moreover, with mild-value of lookahead window size, \textit{POSCAD} achieves a significant reduction in the average request response time.

\subsubsection{Evaluation with Imperfect Prediction}
Prediction errors are inevitable in practice. Hence, we are also interested in the robustness of POSCAD against such errors. In the following, we vary error rate from $0\%$ to $50\%$.

\textbf{Average response time vs. prediction error rate}: 
Figure \ref{fig_acc} shows the system performance with prediction error rate $r$ ranging from $0\%$ (perfect prediction) to $50\%$ under \textit{POSCAD} with various settings. 
On the one hand, with parameter $V = 10$ and window size $D = 2$, we observe a growth ($6.21$ms) of the average response time as the prediction error rate rises from $0\%$ to $50\%$. 
This suggests that prediction errors lead to an increased request response time. With prediction error rate $r$ fixed as $10\%$ and a prediction window size $D = 2$, 
we see that the response time ($21.26$ms) under $V = 10$ is higher than that ($7.9$ms) under $V = 1$. The result suggests that one can reduce request response time by decreasing the value of parameter $V$. The reason is that a smaller value of $V$ leads to a smaller total queue backlog size which, by Little's theorem \cite{little1961proof}, implies a lower average response time. 
Besides, by fixing the prediction error rate (take $r = 10\%$ as an example), we see that the longer the window size, the shorter the response time. This implies that more future information conduces to response time reduction by up to $50\%$ even under inaccurate prediction. 
On the other hand, we see that higher prediction error rates lead to higher system costs, since the system needs to allocate surplus resources to process mis-predicted requests. Besides, there are no significant costs ($\sim 2\%$) incurred as the value of window size $D$ increases. Meanwhile, increasing the value of $V$ results in a notable reduction of response time. Such results show that the advantage of predictive scheduling mainly lies in response time reduction. Its impact is minor to system costs because of the trade-off between cost reduction and queue stability.

\textbf{Insight:} \textit{POSCAD} is robust against prediction errors. By choosing a smaller value of $V$ and enlarging the prediction window size, \textit{POSCAD} can effectively eliminate the negative effect of prediction errors and shortens request response time.

\section{Related Work} \label{sec: related work}
Regarding \textit{switch-controller association}, the usual design choice is to make a static switch-controller association \cite{koponen2010onix}\cite{tootoonchian2010hyperflow}. 
However, solutions with static association are often inflexible when dealing with temporal variations of request traffic, thereby inducing workload imbalance across controllers and increased request processing latency. 
To mitigate such issues, Dixit \textit{et al.} proposed an elastic distributed controller architecture with an efficient protocol design for switch migration among controllers \cite{dixit2013towards}. However, the design for \textit{switch-controller association} still remained unresolved. 
Later, Krishnamurthy \textit{et al.} took a further step by formulating the controller association problem as an integer linear problem with prohibitively high computational complexity \cite{krishnamurthy2014pratyaastha}. 
A local search algorithm was proposed to find the best possible association within a given time limit (\textit{e.g.}, 30 seconds). Wang \textit{et al.} modeled the controller as an $M/M/1$ queueing system \cite{wang2016dynamic}. By formulating the association problem with a steady-state objective function as a many-to-one stable matching problem with transfers, they developed a novel two-phase algorithm that connects stable matching to utility-based game theoretic solutions, \textit{i.e.}, coalition formation game with Nash stable solutions. 
Later, they extended the problem with an aim to minimize the long-term costs in SDN systems \cite{wang2017efficient}. By decomposing it into a series of per-time-slot controller assignment sub-problems, Wang \textit{et al.} applied receding horizon control techniques to solve the problem. In parallel, Filali \textit{et al.} \cite{filali2018sdn} formulated the problem as an one-to-many matching game, then developed another matching-based algorithm that achieves load-balancing by assigning minimum quota of workload to each controller. 
Lyu \textit{et al.} \cite{lyu2018multi} presented an adaptive decentralized approach for joint switch-controller association and controller activation with periodic on-off control to save operational costs. 
Regarding control devolution, most works have focused on static delegation of certain network functions to switches \cite{curtis2011devoflow} \cite{hassas2012kandoo} \cite{zheng2015lazyctrl}. Some recent works \cite{yang2016focus} have proposed more flexible devolution schemes based on the distribution of network states or workloads on controllers in real time.

Compared to existing works, the focus of our work is on the joint design of dynamic switch-controller association and control devolution with performance analysis, while investigating the benefits of predictive scheduling in SDN systems. Particularly, without prediction, our schemes are not dependent on any particular network traffic distributions. However, when adopted in real scenarios with dramatic traffic variations \cite{yi2014building} \cite{liu2016eba}, network traffic prediction would inevitably lead to prediction errors. Nonetheless, our proposed scheme is still robust against such prediction inaccuracy, as illustrated in Section \ref{Section: simulation}.

\section{Conclusion} \label{sec: conclusion}
In this paper, we studied the problem of dynamic switch-controller association and control devolution, and investigated the benefits of predictive scheduling for SDN systems. We proposed  \textit{POSCAD}, an efficient online joint control scheme that solves the problem through a series of online decision making in a distributed manner. 
Our theoretical analysis showed that without prediction, \textit{POSCAD} achieves near-optimal total costs with queue stability guarantee. 
Furthermore, with predictive scheduling, \textit{POSCAD} achieves even better performance with a significant reduction in request response time. We conducted extensive simulations to verify the effectiveness of \textit{POSCAD}. Our results showed that with mild-value of future information, 
\textit{POSCAD} incurs a significant reduction in request response time, even in face of mis-prediction.
{Notably, the focus of this work is the exploration of the \textit{fundamental} limits of the benefits of predictive scheduling in SDN systems. Therefore, we did not consider testbed-based experimental verifications in this work. Nonetheless, it can be an interesting direction for future work.}

%

\ifCLASSOPTIONcaptionsoff
  \newpage
\fi




\end{document}